\documentclass[aps,prfluids,reprint,onecolumn,amsmath,amssymb,superscriptaddress]{revtex4-2}
\usepackage{graphicx}
\usepackage{xcolor,float,mathtools}
\usepackage{dcolumn}
\usepackage{bm}

\begin{document}
	
\title{Fluctuations and power-law scaling of dry, frictionless granular rheology near the hard-particle limit}
\author{A. P. Santos}
\affiliation{Sandia National Laboratories, Albuquerque, NM 87185, USA}
\author{Ishan Srivastava}
\affiliation{Center for Computational Sciences and Engineering, Lawrence Berkeley National Laboratory, Berkeley, CA 94720, USA}
\author{Leonardo E. Silbert}
\affiliation{School of Math, Science and Engineering, Central New Mexico Community College, Albuquerque, NM 87106, USA}
\author{Jeremy B. Lechman}
\email{jblechm@sandia.gov}
\affiliation{Sandia National Laboratories, Albuquerque, NM 87185, USA}
\author{Gary S. Grest}
\affiliation{Sandia National Laboratories, Albuquerque, NM 87185, USA}
\date{\today}
\begin{abstract}
The flow of frictionless granular particles is studied with stress-controlled discrete element modeling simulations for systems varying in size from 300 to 100,000 particles. 
The volume fraction and shear stress ratio $\mu$ are relatively insensitive to system size fo a wide range of inertial numbers $I$.
Second-order effects in strain rate, such as second normal stress differences, require large system sizes to accurately extract meaningful results, notably a non-monotonic dependence in the first normal stress difference with strain rate. 
The first-order rheological response represented by the $\mu(I)$ relationship works well at describing the lower-order aspects of the rheology, except near the quasi-static limit of these stress-controlled flows.
The pressure is varied over five decades, and a pressure dependence of the coordination number is observed, which is not captured by the inertial number.
Large fluctuations observed for small systems $N\le$ 1,000 near the quasi-static limit can lead to arrest of flow resulting in challenges to fitting the data to rheological relationships. 
The inertial number is also insufficient for capturing the pressure-dependent behavior of property fluctuations.
Fluctuations in the flow and microstructural properties are measured in both the quasi-static and inertial regimes, including shear stress, pressure, strain rate, normal stress differences, volume fraction, coordination number and contact fabric anisotropy.  
The fluctuations in flow properties scale self-similarly with pressure and system size.
A transition in the scaling of fluctuations of stress properties and contact fabric anisotropy are measured and proposed as a quantitative identification of the transition from inertial to quasi-static flow.
\end{abstract}
\maketitle
\section{Introduction}\label{sec:intro}
Granular particles with frictionless inter-particle contacts display an effective macroscopic friction and resistance to flow. 
Once the material overcomes the yield stress and flows, it dilates and the shear stress increases as strain rate increases. 
The $\mu(I)$ rheological model for dense, inertial, steady state flows of granular materials has emerged as an accurate description of granular rheology~\cite{Midi2004,Jop2006}. 
The model assumes that in the bulk limit the shear stress $\tau$ to pressure $P$ ratio, or shear stress ratio $\mu = \tau/P$ and the volume fraction $\phi$ vary monotonically with respect to the dimensionless flow rate, inertial number
\begin{eqnarray}
	I=\frac{\dot{\gamma}\bar{d}}{2}\sqrt{\frac{\rho_p}{P}}\label{eqn:I}
\end{eqnarray}
\noindent where $\dot{\gamma}$ is the strain rate, $\bar{d}$ is the average particle diameter and $\rho_p$ is the particle density. The $\mu(I)$-rheology applies quite generally across different flow geometries, including flows down an incline plane~\cite{Jop2006} and rotating drums~\cite{Renouf2005}, and suspension flows where frictional contacts dominate~\cite{Huang2005}. 
Naturally, the goal is to develop a robust continuum description that can be used as a predictive tool for a wide variety of natural and technological processes, including scale-up.
Further development of such continuum descriptions include wall effects\cite{JOP2005}, higher order rheological effects\cite{Srivastava2021}, fluctuations and non-local effects~\cite{Henann2013}.

However, the reduction of the shear and strain rate tensors to $\mu$ and $I$, respectively, loses information that is important for distinguishing many rheological behaviors.
For example, scalar models such as $\mu(I)$ do not explain anomalous stress profiles in cylindrical Couette flow~\cite{Mehandia2012} and negative rod climbing in rotating-rod flow~\cite{Boyer2011}. 
The lack of coaxiality between principal directions of stress and strain rate tensors in viscometric flows~\cite{Alam2003,Depken2007,Weinhart2013,Seto2018} contributes to those effects.
Srivastava \textit{et al}.~\cite{Srivastava2021} developed a second-order rheological model that does not assume coaxiality of stress and strain rate tensors, which is important for capturing the role of inter-particle friction on granular rheology.
Tensorial expressions of shear and strain rate are also important for describing the influence of loading geometries, as shown by Clemmer \textit{et al}.~\cite{Clemmer2021b} in irrotational loading geometries of granular flows, where Drucker-Prager~\cite{Drucker1952} type models can be insufficient.
A goal of this paper is to understand the effect of pressure and system size $N$ on the tensorial second-order rheological model~\cite{Srivastava2021}, without the added contributions due to frictional contacts or suspensions.

In addition to bulk rheological models, non-local models have been developed to describe boundary and finite-size effects.
Non-locality can be described as fluctuations in one area of the material inducing change in another area~\cite{Pouliquen2009}, and is often introduced through a granular fluidity field.  
Fluctuations in microscopic variables, such as the stress~\cite{Pouliquen2009}, strain rate~\cite{Jop2012}, particle velocity~\cite{Zhang2017f}, and force network fluctuations~\cite{Radjai2002,Thomas2019} have been used to characterize granular fluidity in non-local models.
Kinetic theories provide explicit connections between fluctuations and higher order rheological properties, such as the connection between anisotropy in the second moment of velocity fluctuations and normal stress differences~\cite{Santos1998,Alam2003}.
Experiments and simulations have shown that the mean velocity fluctuations scale with the inertial number for a variety of flow geometries~\cite{Midi2004,Pouliquen2004,Gaume2020}, and signal particle friction-dependent flow regimes~\cite{Degiuli2016}. 
The ``granular temperature''~\cite{Edwards1989}, defined as the second moment of the velocity, can be used to understand variance in $\mu(I)$ for different flow configurations~\cite{Kim2020a}.
Going beyond the second moment of the velocity has been used to identify the transition from critical and plastic regimes in granular flows~\cite{Woldhuis2015}.
Fluctuations in other properties of granular flow also have equilibrium thermodynamic relations, such as the volume fraction and compressiblity.
Another goal of this paper is to present the $P$ and $N$ scaling of fluctuations of flow and microstructural properties.

Because fluctuations play a crucial role in granular rheology, a careful analyses of their scaling properties, particularly with system size $N$, is crucial. 
Perrin \textit{et al}. experimentally observed that as the height of frictionless granular flows, and thus number of particles, down an incline increases, the critical stress ratio decreases~\cite{Perrin2021}.
The $\mu(I)$ has been fit by power-laws ($\mu(I)\sim I^{\alpha_{\mu}}$) and other forms~\cite{JOP2005}.  
Simulations have been used extensively to study system-size and pressure effects in frictionless granular flows.  
Simulations of 2d and 3d particles under stress- and strain-controlled simple shear have fit power laws to $\mu(I)$ and $\phi(I)$~\cite{Peyneau2008,Kawasaki2015,FavierdeCoulomb2017,Srivastava2019}.
Those fit parameters have power-law dependencies on $N$ and $P$~\cite{Peyneau2008,Kawasaki2015,FavierdeCoulomb2017}, as was also found for shear-jammed systems~\cite{Xu2006,Taboada2006,Hatano2007,Olsson2007,Heussinger2010,Shojaaee2012,Olsson2020}.
Fits to data from simulations, experiments and different configurations have resulted in a range of power-law exponent values~\cite{Peyneau2008,Forterre2008,Trulsson2012,Bouzid2013,Azema2014,DeGiuli2015,Kawasaki2015,Perrin2021} that match well with theoretical predictions~\cite{DeGiuli2015}.
Fitting such power laws require large amounts of robust data.
In this paper, we examine frictionless granular flows for a large range of $N$, $P$ and $I$, and study the effect of $N$ and $P$ on tensorial granular rheology~\cite{Srivastava2021} and its intrinsic fluctuations.

We present stress-controlled simulations, where flow is induced by applying simple shear to the periodic boundaries of systems with $300 \le N \le 100,000$ frictionless monodisperse spherical 3D particles and pressures $10^{-6}\le P \le10^{-2}$.
We explore the role of pressure and system size on $\mu(I)$, $\phi(I)$, normal stress differences (in Section~\ref{subsec:rheo}), and rheological fluctuations (in Section~\ref{subsec:flux}) of steady state flows. 

\section{Methodology}\label{sec:methods} 
The particles are modeled as spheres of finite size using discrete element, particle-based simulations.  
The spheres are purely repulsive and only interact when in contact, through a Hookean spring-dashpot interaction potential without friction. 
Particle diameters $d_i$ are uniformly distributed from $0.9<d_i<1.1$ to prevent crystallization.
The particle density $\rho_p=1.91$ and a mean particle mass $\bar{m}=1$.  
Some simulations were run with $\rho_p=1$ for $N=10^4$ with no observable impact on the measured properties.
The particle spring and damping parameters are set to $k_n = 1.0$ and $\gamma_n = 0.5 \sqrt{k_n/\bar{m}}$ where the energy scale $k_n\bar{d}^2$ is set by the spring constant and diameter. 
Particle parameters $k_n$ and $\gamma_n$ are kept constant, and the pressure $P$ is varied. 
In the absence of gravity, $k_n$ sets the scale of stress. 
Therefore, varying $P$ and keeping $k_n$ constant, is equivalent to varying $k_n$ and keeping $P$ constant~\cite{Silbert2001}.  Pressures presented here are normalized by $\bar{d}/{k_n}$.
Campbell~\cite{Campbell2005} found that quasi-static flows are not sensitive to the coefficient of restitution, and thus the damping parameter $\gamma_n$.

The assumption of linear elastic behavior for inter-particle contacts is reasonably accurate as a model for sufficiently stiff particles at sufficiently low pressure.  
Note that as an upper limit, for example, glass has a yield stress $\sigma_y \approx 70 $ MPa and would be expected to yield/fracture/fragment, deviating significantly from spherical shape, for $P >> 10^{-3}$.  
Simulations here are not limited to $P < 10^{-3}$, but higher pressures offer comparisons to previous work where particle deformation is ignored.
\footnotetext[1]{See Supplemental Material at the end of the article for details on the role of pressure damping and time step, as well as more fluctuation and coordination number data.}
\footnotetext[2]{The internal stress is calculated from the inter-particle forces and kinetic energy $\sigma_{\alpha\beta}=1/V\Sigma_i\left[\Sigma_{j\ne i}1/2r_{\alpha,ij}f_{\,ij}+m_iv_{\alpha,i}v_{\beta,i}\right]$ where $r_{\alpha,ij}$ and $f_{\alpha,ij}$ are the separation distance and force between particles $i$ and $j$ in the $\alpha$ Cartesian direction, and $m_i$ and $v_{\alpha,i}$ are the $i^{\text{th}}$ particle mass and velocity in the $\alpha$ direction.}

Simulations are initialized with particles at random, non-overlapping positions and low volume fraction $\phi_0 = 0.05$ in a cubic box with periodic boundary conditions. 
Initial translational and rotational velocities were set to zero.
The fully periodic three-dimensional box is able to change shape with triclinic deformations to maintain the applied stress tensor~\cite{Srivastava2019,Santos2020}.
The stress-controlled, periodic boundary simulation box models bulk behavior away from walls, thus avoiding wall effects on $\mu(I)$~\cite{JOP2005,Fazelpour2021,Dsouza2021}.
In particular, the Shinoda-Shiga-Mikami~\cite{Shinoda2004} formulation of a barostat was used in the $N\mathbf{P}_{\text{ext}}H$ ensemble to integrate the positions and momenta of the particles and box, where $N$ is the number of particles, $\mathbf{P}_{\text{ext}}$ is the applied external pressure tensor and $H$ is the enthalpy.  
Stress-controlled simple shear flow is simulated by applying an external stress tensor to the box defined as:
\begin{eqnarray}
\boldsymbol{\sigma}_{\text{ext}}= \begin{bmatrix}
\sigma_{\text{ext},xx} & \sigma_{\text{ext},xy} & 0\\
\sigma_{\text{ext},yx} & \sigma_{\text{ext},yy} & 0\\
0 & 0 & \sigma_{\text{ext},zz}
\end{bmatrix}
\label{eqn:stress}`
\end{eqnarray} 
\noindent where $\sigma_{\text{ext},xx}=\sigma_{\text{ext},yy}=\sigma_{\text{ext},zz}=P_{\text{ext}}$, and the shear stress $\sigma_{\text{ext},xy}=\sigma_{\text{ext},yx}=\tau_{\text{ext}}$, with the other off-diagonal stresses are $\sigma_{\text{ext},xz}=\sigma_{\text{ext},yz}=\sigma_{\text{ext},zx}=\sigma_{\text{ext},zy}=0$. 
The strain rate tensor $\mathbf{D}$ and the Cauchy stress tensors are measured from the box deformation. 
Beyond the applied stress tensor, the barostat also requires two input parameters: $P_{\text{damp}}=0.2256\sqrt{\bar{m}/k_n}$ and $f_{\text{drag}}=0.05$. 
The pressure damping $P_{\text{damp}}$ adjusts the how quickly the box responds to pressure fluctuations in order to maintain the applied stresses. 
The value $P_{\text{damp}}=0.2256\sqrt{\bar{m}/k_n}$ was picked so that simulations reached steady state in a relatively short time. 
Values of $P_{\text{damp}}=2.256$ and $P_{\text{damp}}=0.1128$ were also used in simulations of $N=10^4$ and $P=10^{-4}$, $10^{-5}$ and $10^{-6}$. 
Changing $P_{\text{damp}}$ shifts the range of inertial numbers $I$ accessible to these stress-controlled simulations, but does not change the average steady-state behavior of $\mu(I)$ or $\phi(I)$. 
The effect of $P_{\text{damp}}$ on fluctuations is more complicated and is discussed in Section~\ref{subsec:flux} and shown in the Supplemental Material~\cite{Note1}. 

To stabilize these out-of-equilibrium simulations, particularly when in transit to the steady-state, the drag factor $f_{\text{drag}}$ scales the box change acceleration.
The simulation box under steady state flow continually deforms due to the difference between the external applied stress $\boldsymbol{\sigma}_{\text{ext}}$ and the internal measured stress $\boldsymbol{\sigma}$.  The properties presented are calculated using the internal stress~\cite{Note2}.

Simulations were performed using LAMMPS~\cite{StevePlimton1995,Thompson2022} to integrate Newton's second law with the velocity-Verlet integration scheme.
The simulation time step is set to $\delta t=0.02\sqrt{k_n/\bar{m}}$.  Time steps of $0.01\sqrt{k_n/\bar{m}}$ and $0.005\sqrt{k_n/\bar{m}}$ were also run for a range of applied external shear stress ratios $\mu_{\text{ext}}=\tau_{\text{ext}}/P_{\text{ext}}$ and $P_{\text{ext}}$ for $N=10^4$. 
The different time steps did not show a difference in the measured property behavior.  

For each applied shear stress, pressure and system size, 3 realizations of particles are initialized and simulated.  Property uncertainties are calculated from the individual simulations and across the different realizations.  Uncertainties are propagated from block averaging of individual runs~\cite{Flyvbjerg1989} and as the standard deviation from the 3 different simulations over the steady-state region in time.
Steady state flow was determined if the measured properties uncertainties reach a plateau, with respect to the data blocks used in the block averaging~\cite{Flyvbjerg1989}.
Simulations were run at steady-state for at least as long as the transient time leading to steady state.
Total simulation time varied depending on the pressure applied, ranging from 1x$10^6$ to 8x$10^8$ time steps.
A simulation was not used if any of the 3 configurations arrested (arrest is identified if the strain rate is below a critical value $\dot{\gamma} >10^{-9} \sqrt{\bar{m}/k_n}$~\cite{Srivastava2019}) or disobeyed simple shear flow\cite{Note3}.
Multiple pressures $P=10^{-7},10^{-6},10^{-5},10^{-4},10^{-3}$ and $10^{-2}$ and system sizes $N=3$x$10^2, 10^3, 3$x$10^3, 10^4,3$x$10^4$ and $10^5$ were simulated (27 total $P$ and $N$ states and 1728 total simulations). 
A pressure of $P = 10^{-2}$ is above the yield stress of many materials, such as glass, and thus a real material is expected to exhibit different behavior than these non-deformable particles at higher pressures $P > 10^{-3}$.

\footnotetext[3]{Simple shear was quantified as $|\beta| < 0.95$, where $\beta$ is the vorticity parameter from Giusteri and Seto~\cite{Giusteri2018}. Specifically, $\beta = \frac{1}{\dot{\gamma}} \frac{\mathbf{W}:\mathbf{G}}{\mathbf{G}:\mathbf{G}}$, where $\mathbf{G} = \hat{\mathbf{d}}_3\hat{\mathbf{d}}_1-\hat{\mathbf{d}}_1\hat{\mathbf{d}}_3$. 
	And $\hat{\mathbf{d}}_1$, $\hat{\mathbf{d}}_2$ and $\hat{\mathbf{d}}_3$ are the orthonormal eigenvectors of the symmetric strain rate tensor $\mathbf{D}$ in decreasing order of eigenvalues. 
	Specifically, $\hat{\mathbf{d}}_1$ represents the compression and $\hat{\mathbf{d}}_3$ represents the expansion directions in the plan of the shear flow. 
	The vorticity parameter $\beta$ is used to quantify the flow behavior, where $\beta=1$ corresponds to simple shear and $\beta=0$ corresponds to elongational flow. ~\cite{Srivastava2021}
}

Flow properties are defined with a tensorial formulation of a general rheological model developed previously~\cite{Srivastava2021}.  
Applying the tensorial methodology to the stress-controlled flow data, defines the shear stress ratio, first and second normal stress difference ratios to pressure.
The first-order contribution $\mu=\tau/P$ to the flow is calculated as the rotationally invariant shear stress in the system:
\begin{eqnarray}
\mu = \frac{1}{2\dot{\gamma}P} \boldsymbol{\sigma}:\mathbf{D}
\label{eqn:mu1}
\end{eqnarray}
\noindent where $ \boldsymbol{\sigma}$ is the Cauchy stress tensor, $\dot{\gamma} = 1/2|\mathbf{D}|$ is the strain rate, measured from the box deformation, $p = 1/3 \text{tr}(\boldsymbol{\sigma})$ is the measured pressure and $\mathbf{D}$ is the strain rate tensor.

Non-Newtonian fluids, including granular flows, typically have second-order contributions to the stress in shear flow which can be characterized by normal stress differences~\cite{Guazzelli2018}. In the general rheological model from Srivastava \textit{et al}.~\cite{Srivastava2021} the property
\begin{eqnarray}
\frac{N_0}{P} = -\frac{3}{2\dot{\gamma}^2P}  \boldsymbol{\sigma}:\left(\mathbf{D}^2-\frac{\text{tr}\left(\mathbf{D}^2\right)}{3}\mathbf{I}\right)
\label{eqn:mu2}
\end{eqnarray}
\noindent , or second normal stress difference, is the difference between the mean normal stress in the flow plane and normal stress in the vorticity direction. 
The second-order contributions to the flow representing the difference between the two normal stresses in the flow plane is calculated by
\begin{eqnarray}
\frac{N_1}{P} = \frac{1}{2\dot{\gamma}^2P}  \boldsymbol{\sigma}:\left(\mathbf{D}\mathbf{W}-\mathbf{W}\mathbf{D}\right)
\label{eqn:mu3}
\end{eqnarray}
\noindent is the first normal stress difference where the vorticity tensor $\mathbf{W} = \frac{1}{2}(\nabla\mathbf{v}-\nabla\mathbf{v}^\text{T})$, . 
For homogeneous simple shear stress flow, the second and first normal stress differences can equivalently be defined as $N_0/P = (2\sigma_{zz}-\sigma_{yy}-\sigma_{xx})/2P$ and $N_1/P = (\sigma_{yy}-\sigma_{xx})/P$, respectively~\cite{Seto2018}.

\section{Results}\label{sec:results}
\subsection{Rheology}\label{subsec:rheo} 
Experimental, industrial and natural granular systems have a range of system sizes.
A benefit of granular systems modeling is that it is tractable and practical to simulate real processes where different boundaries impact rheology, such as rotating drums and split-bottom Couette cells, partially because real system sizes are tractable for simulations of spherical discrete-element particles.
System size plays a role in different geometries~\cite{Fenistein2006,Perrin2021}, and therefore it is important to understand system size $N$ and pressure $P$ effects in bulk-like rheology which exists in most flows.
Bulk-like simple shear flow with periodic boundaries in DEM simulations are presented in this section.

\begin{figure}
	\includegraphics[width=0.9\columnwidth]{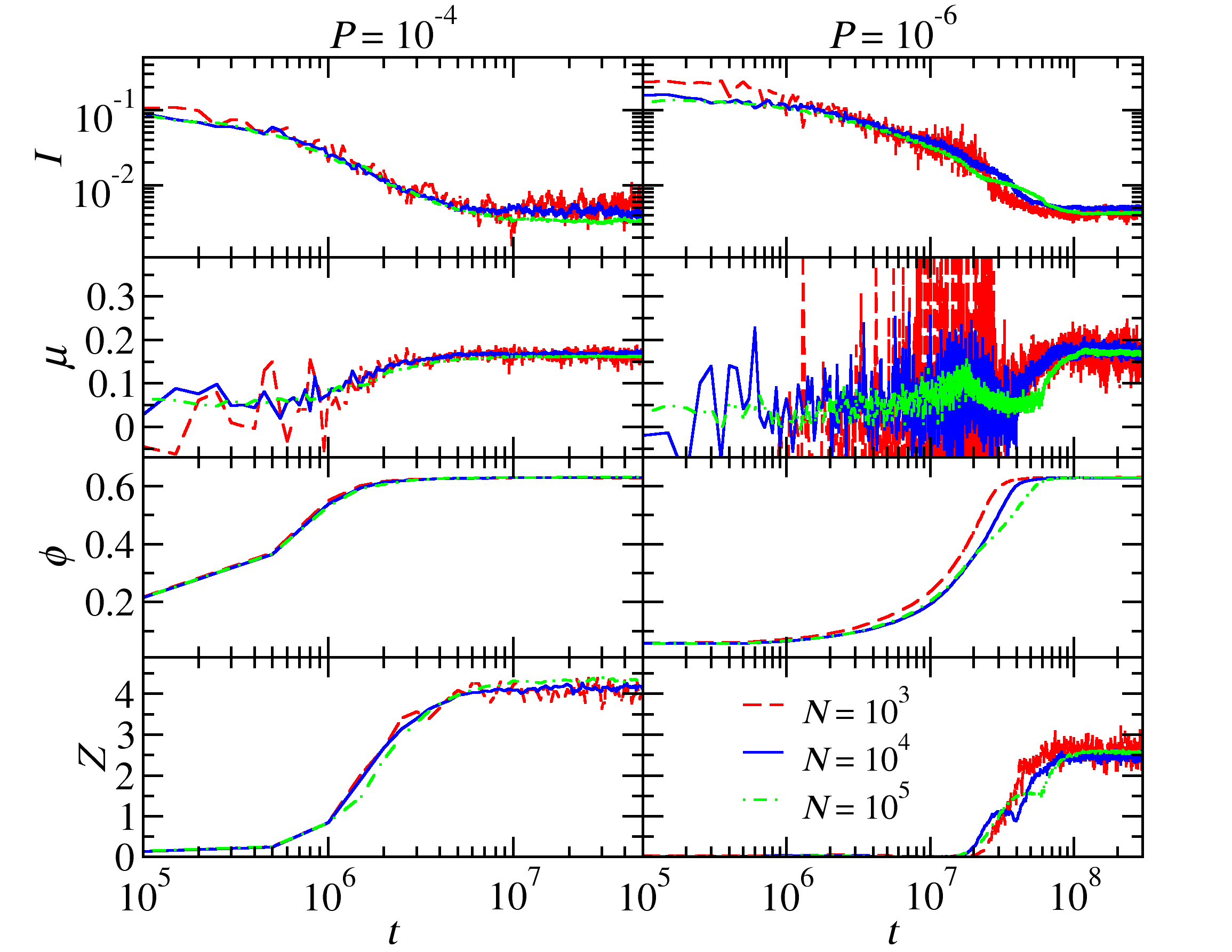}
	\caption{Inertial number $I$, measured stress ratio $\mu$, volume fraction $\phi$, and coordination number $Z$, as a function of time $t$, for three system sizes $N$. Two pressures are shown $P=10^{-4}$ (left-hand panels) and $P=10^{-6}$ (right-hand panels). Different applied shear-stress ratio are shown for each $P$ and $N$, at similar $I\simeq 5$x$10^{-3}$: $\mu_{\text{ext}}=0.2, 0.193, 0.18$ for $P=10^{-4}$ and $\mu_{\text{ext}}=0.4, 0.445, 0.4$ for $P=10^{-6}$ for $N=10^{3}, 10^{4}, 10^{5}$ respectively.
	\label{fig:timeseries}}
\end{figure}

The $P$ and $N$ impact microstructural and flow properties in stress-controlled simulations of granular flow. 
Figure~\ref{fig:timeseries} illustrates the time progression of microstructural and flow properties. 
The inertial number $I$, shear stress ratio $\mu$, volume fraction $\phi$ and coordination number $Z$ are plotted against time.
Each property is shown for two different pressures, $P=10^{-4}$ (left panels) and $P=10^{-6}$ (right panels), and for different system sizes, shown as different colors. 
Pressure nor system size impact the steady-state average $\mu$ and $\phi$ at the same $I$ for the two pressures. 
However, $P$ and $N$ impact the average coordination number $Z$ and fluctuations of all properties, and those impacts are analyzed and discussed in Section~\ref{subsec:flux}.

Figure~\ref{fig:timeseries} shows how stress-controlled simulations of granular flow methodology behave. 
Early times show the transition from a very dilute ($\phi=0.05$) gas-like starting state to a flowing dense, quasi-static regime.
Decreasing the pressure, increases the time to reach steady state due to the the pressure control protocol~\cite{Santos2020}. 
The transient process to steady state is studied elsewhere~\cite{Srivastava2019} and is not the subject of this study.
As discussed in Section~\ref{sec:methods}, an external pressure and shear stress are applied in these simulations.  
The external pressure and shear stress ratio are not equal to the steady-state measured $P$ and $\mu$. 
Figure~\ref{fig:timeseries} shows that $\mu<\mu_{\text{ext}}$, and that as $P$ decreases, more $\mu_{\text{ext}}$ is required to reach the same $I$.
The simulation box and volume fraction fluctuate around the steady-state value, unlike volume-controlled simulations.

\begin{figure}[]
	\includegraphics[width=0.8\columnwidth]{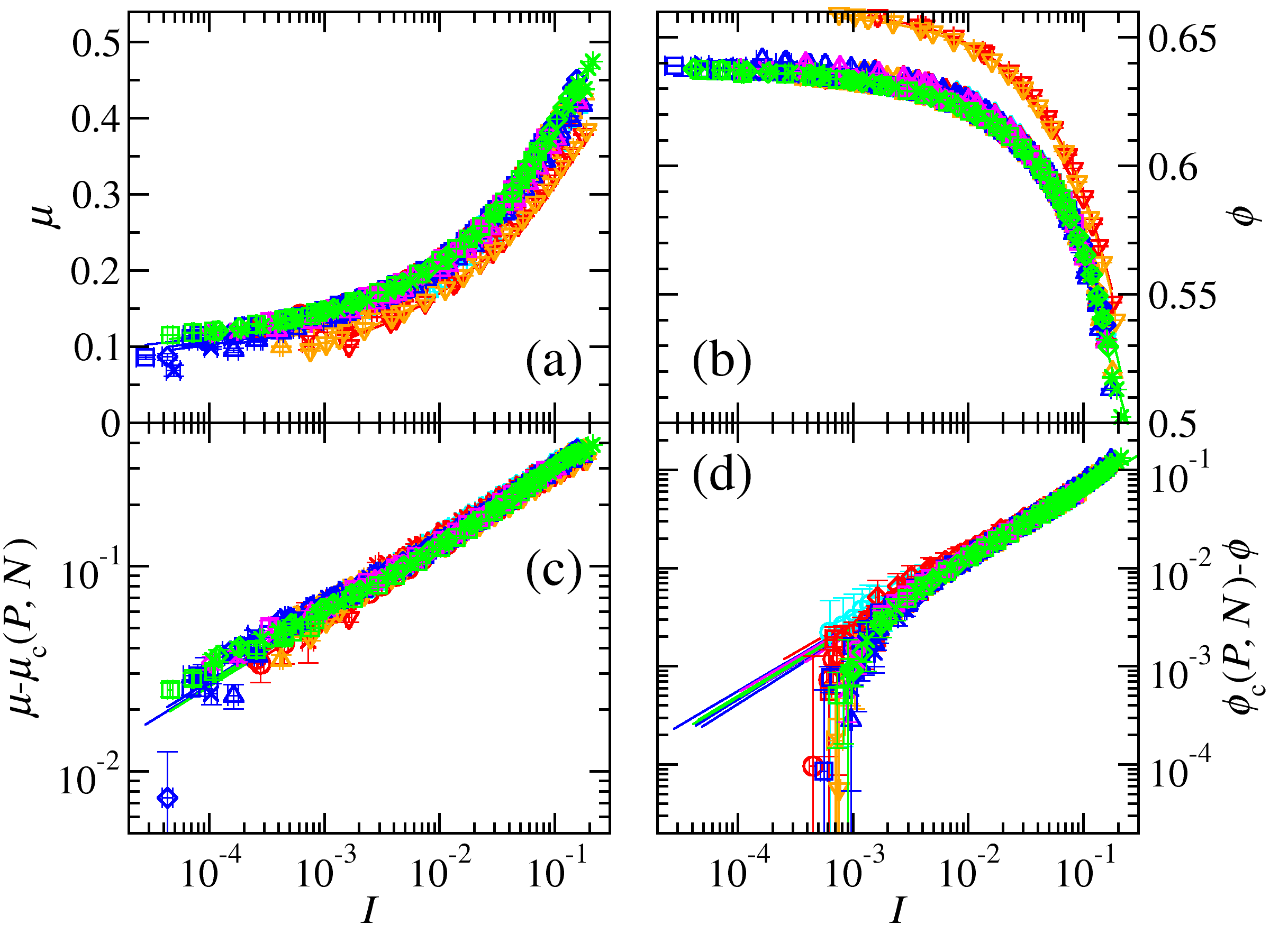}
	\caption{
	The (a) stress ratio $\mu=\tau/P$  and (b) volume fraction $\phi$ as a function of the inertial number.  Subtracting the critical stress ratio $\mu_c$ and volume fraction $\phi_c$, which is calculated from a power-law fit in the limit of $I\to0$, gives (c) $\mu-\mu_c(P,N)$ and (d) $\phi-\phi_c(P,N)$, and illustrates the sensitivity of the power law exponent.  
	Each symbol represents simulation data averaged over three runs at different applied stress, pressure (symbol shape) and system size (symbol color). Simulations were run with $N = 3$x$10^2$ (cyan), $10^3$ (red),  $3$x$10^3$ (orange), $10^4$ (blue), $3$x$10^4$ (magenta) and $10^5$ (green), for $P=10^{-7}$ (circles), $10^{-6}$ (squares), $10^{-5}$ (diamonds), $10^{-4}$ (crosses), $10^{-3}$ (triangles), $10^{-2}$ (inverted triangles).
\label{fig:muphi-I}
}
\end{figure}

\begin{figure}[]
	\includegraphics[width=0.8\columnwidth]{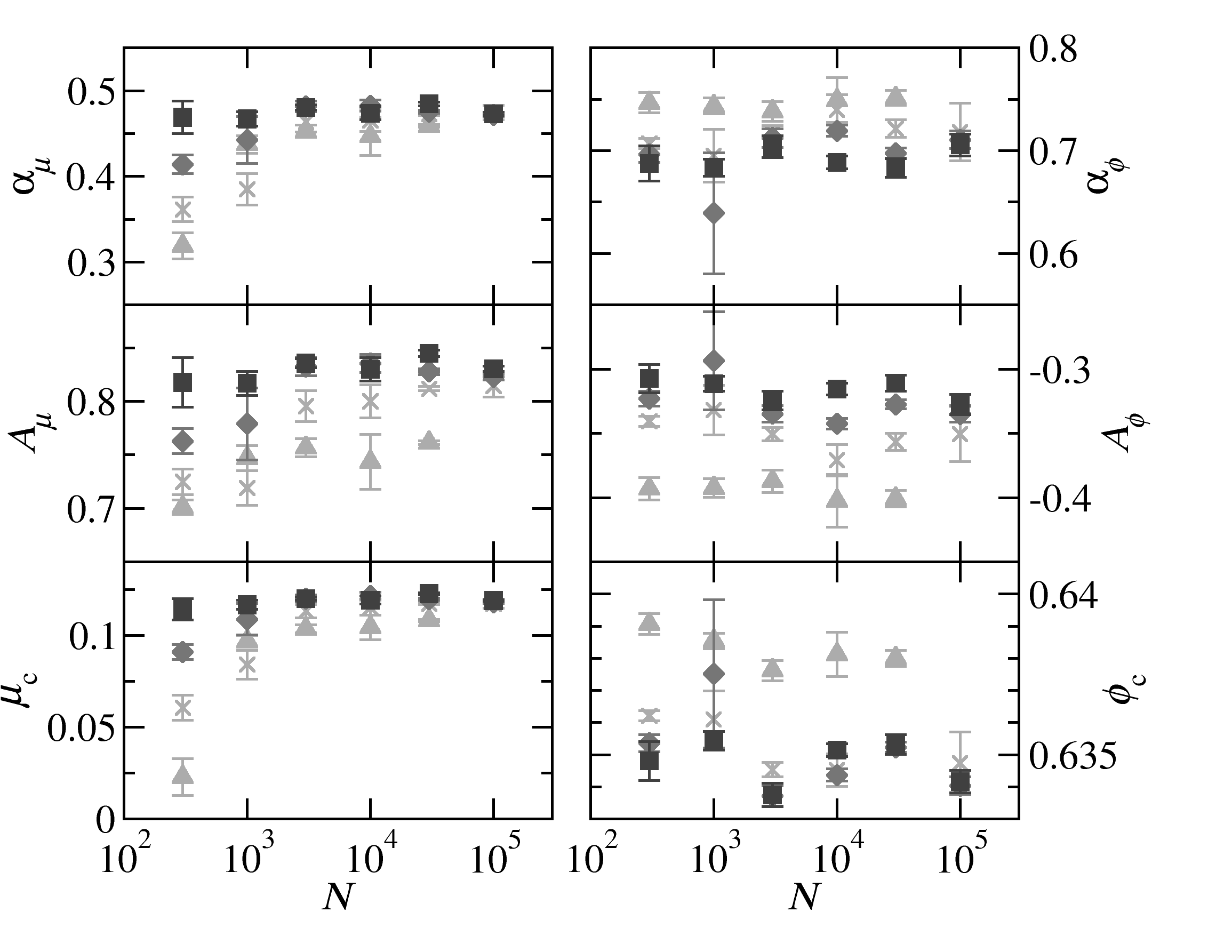}
	\caption{Power-law fit parameters to the stress ratio $\mu(I)$ (left three panels) and volume fraction $\phi(I)$ (right three panels) using Eqs.~\ref{eqn:muI} and~\ref{eqn:phiI}, respectively. The exponents $\alpha_{\mu},\alpha_{\phi}$, pre-factors $A_{\mu},A_{\phi}$ and critical values $\mu_c,\phi_c$ are shown as a function of the number of particles $N$. Different applied pressures are shown as different symbols $P=10^{-7}$ (circles), $P=10^{-6}$ (squares), $P=10^{-5}$ (diamonds), $P=10^{-4}$ (crosses) and $P=10^{-3}$ (triangles).}
	\label{fig:muphifits}
\end{figure}

Within the steady flow regime, all the systems studied here - spanning system sizes and applied pressures - exhibit the expected $\mu(I)$ rheology, as shown in Fig.~\ref{fig:muphi-I}. 
More specifically, Figs.~\ref{fig:muphi-I}a and~\ref{fig:muphi-I}b show that for lower pressures ($P<10^{-3}$), $\mu(I)$ and $\phi(I)$ give very similar results, regardless of pressure or system size. 
Whereas higher pressures, $P\ge10^{-3}$ have a noticeable shift in value, and correspond with particle stiffness values sufficient to model inter-particles linear elastic contact behavior discussed in the methodology.  A power-law, of the form
\begin{eqnarray}
\mu(I) = \mu_c + A_{\mu}I^{\alpha_{\mu}}\label{eqn:muI}\\
\phi(I) = \phi_c - A_{\phi}I^{\alpha_{\phi}}\label{eqn:phiI}
\end{eqnarray}
is fit to the rheology data, and are drawn as the lines in Fig.~\ref{fig:muphi-I}. 
From the power-law fits critical values, $\mu_{c}$ and $\phi_{c}$, are extracted individually for each of the system size and pressure as the value corresponding to the limit: $\mu_{c} \equiv \lim\limits_{I\to0}\mu$ and $\phi_{c} \equiv \lim\limits_{I\to0} \phi$, respectively. 
The shifted data,  $\mu - \mu_c(P,N)$ and $\phi - \phi_c(P,N)$ are shown in Figs.~\ref{fig:muphi-I}c and~\ref{fig:muphi-I}d, demonstrate that all the data are fit well by Eqs.~\ref{eqn:muI} and~\ref{eqn:phiI}.
The shifted $\mu$ and $\phi$ data also demonstrate that the critical, low strain values $\mu_c$ and $\phi_c$ account for the impact of higher pressures $P\ge10^{-3}$.
The fit values are plotted in Fig.~\ref{fig:muphifits}. 

As visually apparent from Fig.~\ref{fig:muphi-I}, most of the fit values in Fig.~\ref{fig:muphifits} are system-size independent with some important exceptions.  
Higher pressure $P\ge10^{-3}$ fit values do show a statistically significant dependence on $N$ in $\mu_c$, $\alpha_{\mu}$ and $\phi_c$, as observed by Peyneau \textit{et al}. for $\mu_c$ and $\phi_c$~\cite{Peyneau2008}.
The other fit parameters, $A_{\mu}$, $A_\phi$ and $\alpha_{\phi}$, depend on $P$ but not $N$.
The exponents $\alpha_{\mu}$ and $\alpha_{\phi}$ are not constant with $N$ and $P$ when allowed to vary. 
Good quality fits ($R^2>0.986$) and discernible change in the other fit value trends can be attained by setting $\alpha_{\mu}(P,N)=0.35$, as proposed by DeGiuli \textit{et al}.~\cite{DeGiuli2015}.
However, better fits are attained when $\alpha_{\mu}(P,N)$ is allowed to vary, as shown in Fig.~\ref{fig:muphifits}.
In the Supplemental Material~\cite{Note1}, the same data in Fig.~\ref{fig:muphifits} is shown with $P$ on the $x$-axis, to aid in understanding the impact of $N$ and $P$.

The fit values in Fig.~\ref{fig:muphifits} can be sensitive to the range of inertial numbers, and it is important to collect data over the range of inertial numbers.  
Only inertial numbers $I$ that were available to all $P$ and $N$ runs (6x$10^{-3}< I < 6$x$10^{-2}$) are used in the fitting procedure so that comparisons, although the range including all simulations is 4 orders of magnitude (3x$10^{-5}< I < 2$x$10^{-1}$). 
Therefore, a wide range of inertial numbers were collected to ensure the fits are representative of the quasi-static and inertial flow regimes.
Collecting data for low inertial numbers requires longer simulations because of the larger fluctuations and longer transient times, up to $\sim2$x times longer.
In these stress-controlled simulations the inertial number is limited on the low end by the transition to stick-slip and arrest behavior.  
For higher inertial numbers the flow becomes more dilute with fewer contacts which contribute to the internal stress. 
Larger strain rates and inertial numbers are thus not accessible to stress-controlled simulations because the flow is driven by the difference in external and internal stress.

\begin{figure}[]
	\includegraphics[width=0.7\columnwidth]{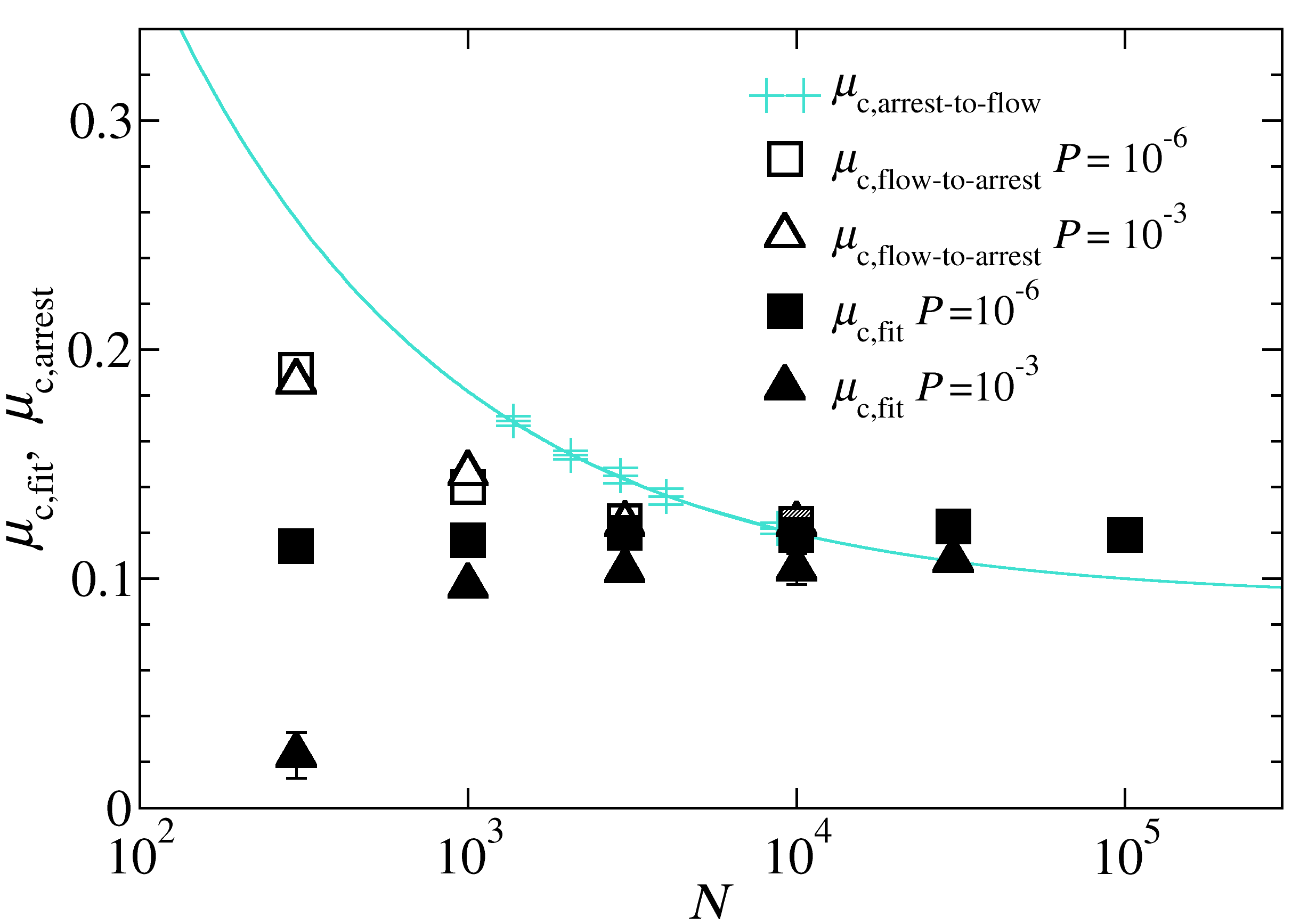}
	\caption{Critical stress ratio calculated by fitting Eq.~\ref{eqn:muI} to $\mu(I)$, $\mu_{\text{c,fit}}$, and calculated as the maximum applied stress where arrest was observed, $\mu_{\text{c,arrest}}$. Arrest was not observed for $N>3$x$10^3$, and, therefore, $\mu_{\text{c,arrest}}$ for $N>3$x$10^3$ are not presented.  The shaded point at $N=10^4$ is from Srivastava \textit{et al}.~\cite{Srivastava2019}, which is a rigorous estimate of the value. Cyan symbols are from static, stress-controlled simulations going from arrest to flow, and the cyan line is a fit, $0.091+2.87N^{-1/2}$~\cite{Peyneau2008}.}
	\label{fig:fitting}
\end{figure}

Although the arresting flows observed in stress-controlled simulations limit the fitting range of $I$, the method gives an estimate of the flow-to arrest transition system size dependence. 
Figure~\ref{fig:fitting} shows the maximum applied stress where a flow-to-arrest transition was observed $\mu_{\text{c,arrest}}$ (open symbols) as a function of number of particles for two pressures.  
The critical stress ratios extracted from the power-law fit to the data $\mu_{\text{c,fit}}$ (as shown with closed symbols) are another measure of arrest.
For large systems, $\mu_{\text{c,arrest}}$ and $\mu_{\text{c,fit}}$ agree.  
As the system size decreases, the values of $\mu_{\text{c}}$ for different methods diverge, specifically $\mu_{\text{c,arrest}}$ increases and $\mu_{\text{c,fit}}$ decreases.  

Arrest is naturally observed in these bulk-like stress-controlled simulations with periodic boundaries.  
As the system size decreases, so does the length of a force chain needed to span the simulation box, and the transition from flow to arrest occurs more frequently and at higher inertial numbers. 
The system size dependence of $\mu_{\text{c,arrest}}$ affects the accessible strain rates and $I$.  
At low system sizes, there is a smaller range of $I$ to fit, which typically leads to a lower $\mu_{\text{c,fit}}$.

Figure~\ref{fig:fitting} also shows that $\mu_{\text{c,arrest}}$ is not pressure dependent, unlike $\mu_{\text{c,fit}}$.
The disagreement between $\mu_{\text{c,arrest}}$ and $\mu_{\text{c,fit}}$ for low $N$ illustrates the importance of system size for characterizing and fitting the $\mu(I)$ rheology in the quasi-static limit.
The fit to data with $N=3$x$10^2$ and $P=10^{-3}$, for example, predicts that applying a stress ratio greater than 0.03 is sufficient to keep the granular material flowing, however the very small system arrests quickly in the simulated realizations.  
The $\mu_{\text{c,arrest}}(N)\simeq N^{-1/2}$ dependence has been observed previously by Peyneau and Roux, and is plotted as the line in Fig.~\ref{fig:fitting}~\cite{Peyneau2008}.
Peyneau and Roux used static stress-controlled simulations, starting from zero shear stress and incrementally increasing the stress until steady-state flow was observed. 
The hysteresis of flow-to-arrest and arrest-to-flow could explain why $\mu_{\text{c,arrest-to-flow}}(N)>\mu_{c,flow-to-arrest}(N)$. 
Hysteresis of the critical stress to flow has been observed in other flow geometries, such as the difference between $\theta_{\text{start}}$ and $\theta_{\text{repose}}$ for flow down an incline\cite{Silbert2001,Silbert2005}.
All the simulated $\mu_{\text{c,\text{arrest}}}$ data presented here represents the flow-to-arrest transition $\mu_{\text{c,flow-to-arrest}}$, except for that from Peyneau and Roux~\cite{Peyneau2008} which is $\mu_{\text{c,arrest-to-flow}}$. 

\begin{figure}[]
\includegraphics[width=0.6\columnwidth]{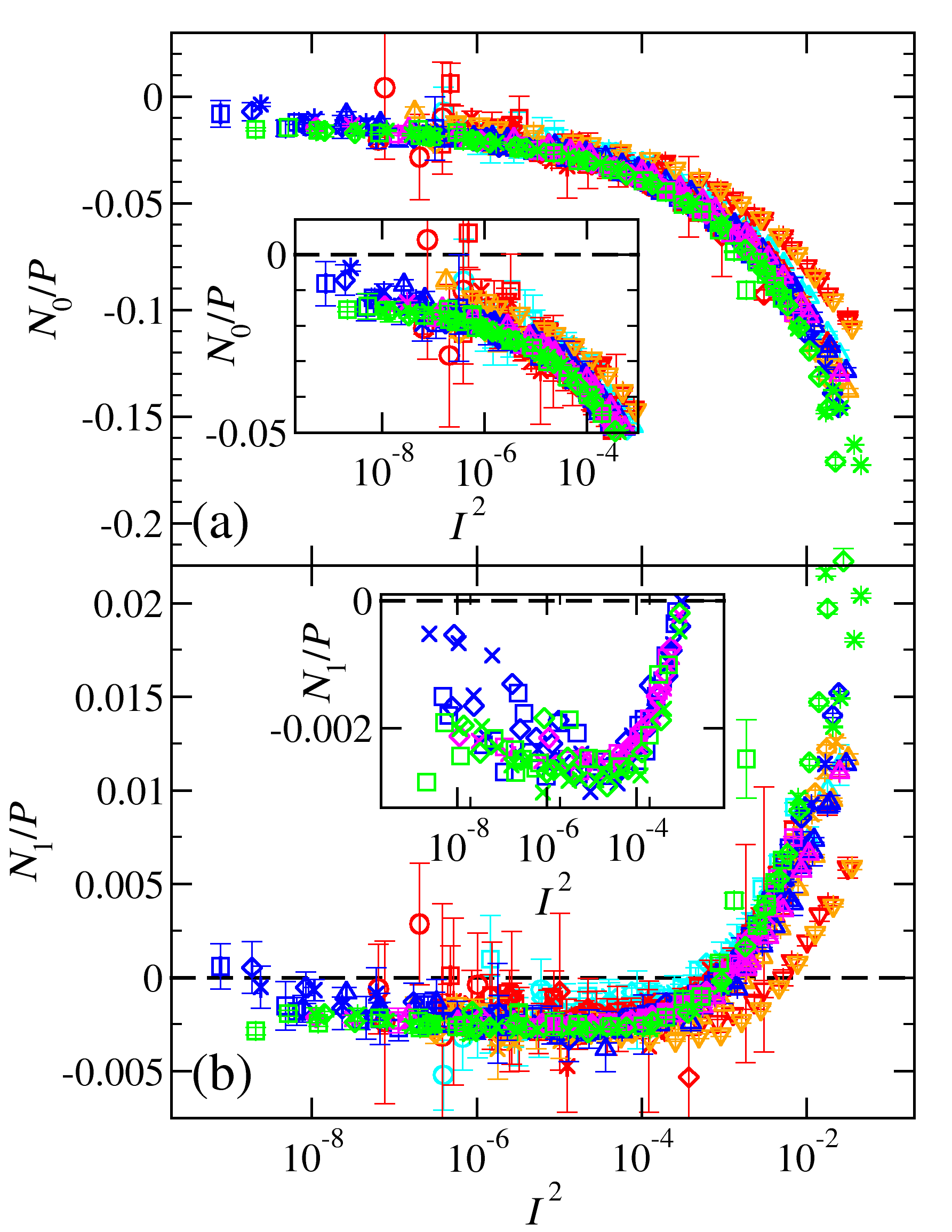}
\caption{(a) Second $N_0/P$ and (b) first $N_1/P$ normal stress difference ratios normalized by pressure (see Eqs.~\ref{eqn:mu2} and~\ref{eqn:mu3}) as a function of $I^2$. System sizes and pressures are the same as in Fig.~\ref{fig:muphi-I}. The inset of (b) shows a close-up near the quasi-static limit of the three larger system sizes ($N=10^4$, 3x$10^4$ and $10^5$).}
\label{fig:N0N1}
\end{figure}

Like the critical stress required to flow granular material, the non-zero first and second normal stress differences distinguish granular flows from simple Newtonian fluids~\cite{Srivastava2021}.  
The first $N_1/P$ and second $N_0/P$ stress differences are shown in Figs.~\ref{fig:N0N1}b and~\ref{fig:N0N1}a, respectively.  
The second normal stress difference is negative in all flows and approaches a non-zero plateau, as the inset of Fig.~\ref{fig:N0N1}a shows. 
The negative values of $N_0/P$ is due to larger normal stress, and number of contacts, in the flow plane as compared to the neutral, vorticity direction, and that difference decreases as $I\to0$.
Although it is expected that $N_0\to 0$ as $I\to0$ for frictionless granular flows~\cite{Srivastava2021}.

As shown in Fig.~\ref{fig:N0N1}b, $N_1/P$ changes from positive to negative as the flow slows, which has been observed previously in experiments~\cite{Couturier2011} and simulations~\cite{Alam2003,Weinhart2013,Seto2018,Srivastava2021} . 
The behavior as $I\to0$ however is debated. 

Both $N_0/P$ and $N_1/P$ are more pressure dependent than $\mu$ and $\phi$, and large system sizes demonstrate important features.
The inset of Fig.~\ref{fig:N0N1}b shows that in the dense-flow regime $N_1/P$ has a minimum.  
The minimum is most convincing and statistically certain for the larger system sizes $N\ge10^4$, which demonstrates the importance of large system sizes for measuring higher-order flow properties. 
The minimum has also been observed in other flow conditions, including flow-down-incline in two-dimensions~\cite{Silbert2001}.  
One explanation in these simple shear simulations is that the misalignment between the fabric and strain-rate tensors~\cite{Seto2018,Srivastava2008}.
The power-law fit used for $\mu(I)$ does not describe $N_1/P(I)$ at low inertial numbers because of the minimum.

Seto \textit{et al}. observed that dense suspensions also have non-zero plateaus in $N_1/P$, and that the plateau goes to zero as the particle spring constant increases (100x increase in spring constant lead to 10x decrease in the  $N_1/P$ plateau)~\cite{Seto2018}. 
In the presented data, there is no pressure dependence in the low $I$ regime across 5 orders of magnitude.
The $N_0/P$ and $N_1/P$ for other timesteps are shown in the Supplemental Material~\cite{Note1}.
Because decreasing the timestep does not impact the mean values of either $N_0/P$ or $N_1/P$, it is assured that the timestep $\delta t=0.02 \sqrt{k_n/\bar{m}}$ is not too large to capture the small fluctuations in stress at low inertial numbers (a concern posed by Seto \textit{et al}.~\cite{Seto2018}), at least for these stress-controlled simulations of dry frictionless particles.  

\subsection{Fluctuations}\label{subsec:flux} 
The transition from quasi-static to inertial granular flow is gradual in the average values of $\mu(I), \phi(I), N_1(I)$ and $N_0(I)$.
Fluctuations about those averages however have been very useful for signaling transitions.
For example, in the approach to jamming, velocity~\cite{Lootens2003} and viscosity~\cite{Hoffman1972} fluctuations become discontinuous. 
Velocity fluctuations can also signal the transition from critical and plastic regimes in granular flows~\cite{Woldhuis2015}.
Fluctuations can characterize non-local effects~\cite{Henann2013} and comparisons between geometries~\cite{Kim2020a}.
In particular, particle stiffness, or pressure, impacts strain rate and kinetic energy fluctuations in granular flow and leads to different flow regimes~\cite{FavierdeCoulomb2017}.
In the previous section the range of accessible $I$ values was limited in low system sizes, because of the large fluctuations that lead to arrest.
It is therefore important to study the impact of system size $N$ and pressure $P$ on fluctuations.

\begin{figure*}
	\includegraphics[width=\linewidth]{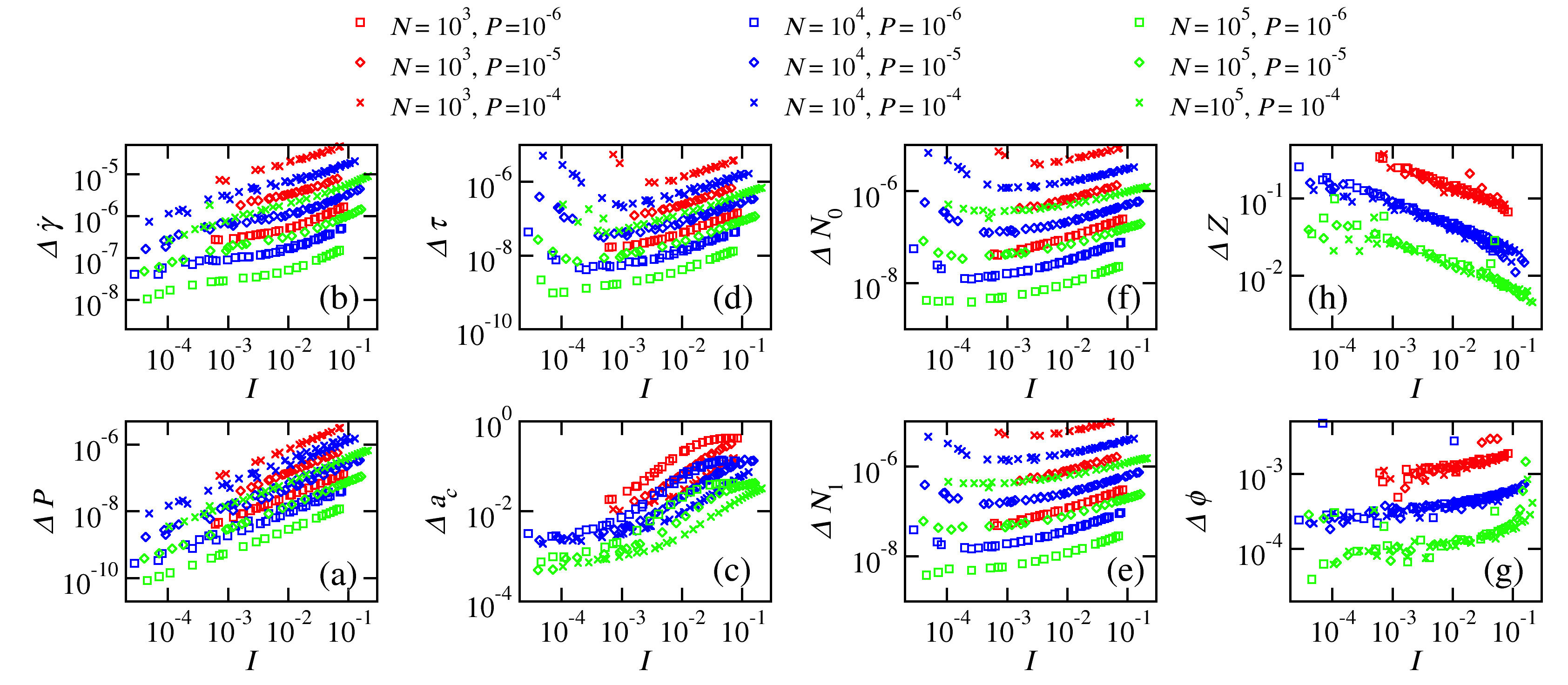}
	\caption{The variance $\Delta$ of steady-state fluctuations of the following parameters as a function of the measured steady-state inertial number $I$: (a) pressure $P$, (b) strain rate $\dot{\gamma}$, (c) contact fabric anisotropy $a_c$, (d) shear stress $\tau$, (e) first normal stress difference $N_1$, (f) second normal stress difference $N_0$, (g) volume fraction $\phi$, and (h) coordination number $Z$. The colors and symbols represent different $N$ and $P$, and are the same as in Fig.~\ref{fig:muphi-I}.}
	\label{fig:var}
\end{figure*}

The time series shown in Fig.~\ref{fig:timeseries} illustrates the fluctuation of kinematic ($I$), mechanical ($\mu$) and microstructural ($\phi$ and $Z$) properties and how they depend on $P$ and $N$ about their mean.
The variance of those properties over the steady flow period quantifies those fluctuations.  
The variance of, for example, the shear stress $\tau$ is defined as $\Delta \tau \equiv \frac{1}{N_{\text{samp}}}\Sigma_{i=1}^{N_{\text{samp}}}\left(\tau(t)-\bar{\tau}\right)$ over the steady state simulation data.  
In addition to $\dot{\gamma}$, $\tau$, $\phi$ and $Z$, we also analyze fluctuations in the stress differences $N_0$, $N_1$ and structural anisotropy of the particle contact network.
The structural anisotropy of the contact network is quantified by the second invariant of the deviatoric contact anisotropy tensor $a_c$. 
The tensor components of $a_c$ in the $i,j$ direction are $a_{c,ij}=\frac{15}{2}R'_{ij}$ and the contact fabric tensor, based on the contact normals $n_i$ of $N_c$ total contacts, is $R'_{ij}=
\frac{1}{N_c}\Sigma_{N_c}n_in_j$~\cite{Radjai2012}.
The variance of flow properties are shown as a function of inertial number in Fig.~\ref{fig:var}. 

The variance of most properties increases with increasing inertial number, including $\Delta P, \Delta \dot{\gamma}, \Delta a_c$ and $\Delta \phi$. 
Whereas for $\Delta Z$, as the material flows faster, the fluctuations decrease.
The  flow properties, $\Delta \tau, \Delta N_1$ and $\Delta N_0$ have a non-monotonic dependence on $I$.
$\Delta \tau, \Delta N_1$ and $\Delta N_0$ behave like the other properties, above a critical $I>I_c$.  
Below that critical $I<I_c$, the fluctuations increase approaching arrest.
As for $\Delta \tau, \Delta N_1$ and $\Delta N_0$, velocity fluctuations of flowing granular materials also grow near jamming~\cite{Lootens2003}.

\begin{figure*}
	\includegraphics[width=\linewidth]{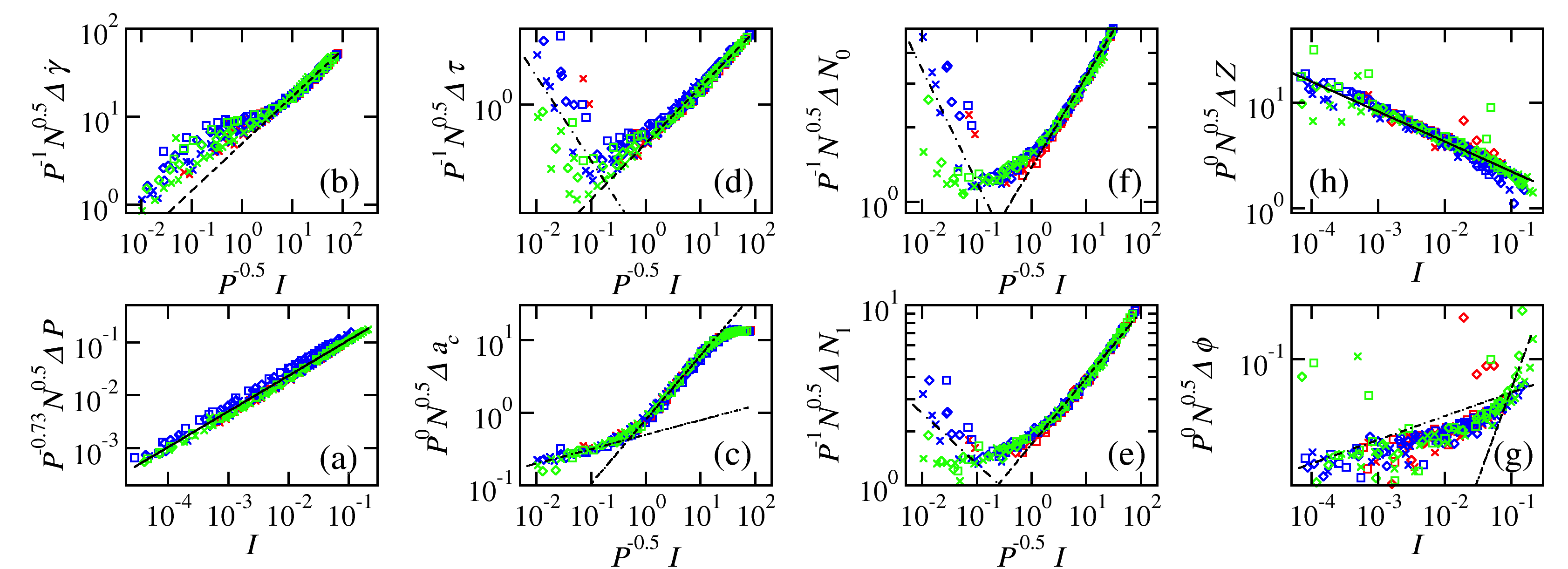}
	\caption{Variance $\Delta$ normalized by $N^a$ and $P^b$ for various properties, which leads to a collapse as a function of the inertial number $I$, except for the stress values, $\tau$, $N_1$, $N_0$ and the fabric anisotropy $a_c$ which are normalized by $P^c$. The normalized variance is fit to a power law over the whole range (solid lines), a low range (dot-dashed lines) and/or a high range (dashed lines) of $I$ or $P^{-0.5}I$. The exponents of the power-law fits drawn as lines are shown in Tab.~\ref{tab:exponents}.}
	\label{fig:varnorm}
\end{figure*}

Fluctuations of flow properties depend on $N$ and $P$, beyond the $I$, unlike the average flow properties.
The $N$ and $P$ dependence of the variance are shown in Fig.~\ref{fig:varnorm}. 
Collapse of all the variance data as a function of strain rate $\dot{\gamma}$ and pressure $P$ is possible with different scalings, as shown in the y-axis label in Fig.~\ref{fig:varnorm}. 
The fluctuations of each property were scaled as
\begin{align}
N^aP^b\Delta &= P^cI = B\dot{\gamma}P^{-0.5}P^c\label{eqn:var}
\end{align}
\noindent where $B=\frac{\bar{d}}{2}\sqrt{\rho_p}$. 
The exponents $a$ and $b$ are applied to the variance, and depend on the property. 
$\Delta\tau(I), \Delta a_c(I), \Delta N_1(I)$ and $\Delta N_0(I)$ have $P$-dependent transitions where the slope changes, and thus $I$ is scaled by $P^c$. 

Power-law fits to the scaled variances are shown in Fig.~\ref{fig:varnorm}. 
Applying a power-law fit to Eq.~\ref{eqn:var} leads to:
\begin{align}
N^aP^b\Delta &\sim \left(P^cI\right)^d\label{eqn:varfit}
\end{align}
\noindent where $d$ is the power-law exponent.  
Solving Eq.~\ref{eqn:varfit} for the variance leads to the following scaling law for kinematic, stress and microstructural property variance:
\begin{align}
\Delta &\sim N^{-a}\dot{\gamma}^dP^{d(c-1/2)-b}\label{eqn:varfitnor}
\end{align}
\noindent and we define the pressure exponent as $e\equiv d(c-1/2)-b$.
The exponents of the fits $d$ and the pressure exponent $e$ are shown in Tab.~\ref{tab:exponents}. 

\begin{table}
	
	\caption{Exponent values used to scale, normalize and fit the the variance of various properties run at different number of particles $N$ and applied pressure as a function of inertial number.}
	\begin{tabular}{c| ccc|cccc }
		\hline
		\hline
		~~~~~~property~~~~~~    & ~~~~~$a$~~~~~ & ~~~~~$b$~~~~~ & ~~~~~$c$~~~~~ & ~~~~~~$d$~~~~~~ & ~~~~~~$e^*$~~~~~~ &  ~~~~~~$d^+$~~~~~~ & ~~~~~~$e^+$~~~~~~  \\
		\hline
		$P$                     & 0.5 & -0.73 & 0 & 0.686$\pm$0.003 & 0.387 & &  \\
		$\dot{\gamma}$ & 0.5 & -1 & -0.5 &0.54$\pm$0.01& 0.73 & &  \\
		$a_c$             & 0.5 & 0 & -0.5 & 0.20$\pm$0.02 & -0.1 & 0.880$\pm$0.008 & -0.88 \\
		$\tau$         & 0.5 & -1 &-0.5 & -0.8$\pm$0.2 & 1.8 & 0.519$\pm$0.005 & 0.481  \\
		$N_1$               & 0.5 & -1 & -0.5 & -0.28$\pm$0.08& 1.28 & 0.367$\pm$0.003& 0.633 \\
		$N_0$               & 0.5 & -0.5 & -0.5 & -0.5$\pm$0.1& 1.5 & 0.369$\pm$0.003& 0.61 \\
		$\phi$              & 0.5 & 0 & 0 & 0.12$\pm$0.02 & -0.06 & 0.58$\pm$0.07  & -0.29  \\
		$Z$                 & 0.5 & 0 & 0 & -0.28$\pm$0.01 & 0.14 & & \\
		\hline
	\end{tabular}
	
	\begin{flushleft}	
		\noindent$^*$The variance-pressure exponent $e=d(c-1/2)-b$.
		\newline\noindent$ ^+$Property variance with different fits at high strain rates.
		\end{flushleft}
	\label{tab:exponents}
\end{table}

For all properties, smaller system sizes have larger fluctuations ($a>0$). 
An exponent of $a=1/2$ is expected from the central limit theorem, and has been seen previously for $\Delta \phi$ and $\Delta \tau/P$ of frictionless granular flows~\cite{Peyneau2008}.
For most of the properties, fluctuations decrease with pressure ($b<0$).
However, the microstructural properties  $\Delta a_c$, $\Delta \phi$ and $\Delta Z$ are $P$-independent ($b=0$)  with respect to the inertial number. 
The magnitude of the fit exponents $|d|$ varies from 0.12 to 0.880 depending on the specific property.

Although it is expected that at a given inertial number the fluctuations decrease either when $P$ increases or $N$ decreases, the transitions of two different slopes in $\Delta \tau, \Delta a_c, \Delta N_1, \Delta N_0$ and $\Delta \phi$ are surprising. 
The transitions in the normalized fluctuations of $\tau, N_1, N_0$ and $a_c$, measured for $P^{-0.5}I$ at the intersection of two fits, are statistically consistent, $P^{-0.5}I$ = $0.2\pm0.1$, $0.3\pm0.1$, $0.2\pm0.1$ and $0.5\pm0.2$, respectively.  
No such transition is observed in the average steady-state values of $\tau, N_1, N_0$ or $a_c$, which are essentially $N$- and $P$-independent and have the same $I$ dependence, as seen in Figs.~\ref{fig:muphi-I} and~\ref{fig:N0N1}. 
Thomas \textit{et al}.~\cite{Thomas2019} saw a transition in the slope of the force network fluctuation rate at a similar inertial number $I\simeq0.06$ in granular flow-down-an-incline experiments.
This change in scaling at low $P^{-0.5}I$, which occurs for all the mechanical property fluctuations could be a quantitative measure of the transition from quasi-static to inertial flows.  
The transitions in the scaling $\Delta \tau$ and $\Delta a_c$ are similar, because the contact fabric is the primary support for the shear stress~\cite{Srivastava2020}.

The transition in scaling of the normalized $\Delta \tau$ and $\Delta a_c$ at $P^{0.5}I=0.2\pm0.1$ depends not only on pressure but pressure damping $P_{\text{damp}}$.
Pressure damping also impacts the fluctuations of $\Delta P$ and $\Delta\dot{\gamma}$.
Fluctuations of properties which are sensitive to the pressure are expected to be sensitive to numerical pressure control parameters in the stress-controlled simulation method.
See the Supplemental Material~\cite{Note1} for figures showing the impact of $P_{\text{damp}}$ on fluctuations in systems with $N=10^4$ particles with $P = 10^{-4}$, $10^{-5}$ and $10^{-6}$ with $P_{\text{damp}} =$ 2.256, 0.2256 and 0.1128. 

The Supplemental Material~\cite{Note1} also includes the relative variance, normalized by the absolute mean $\bar{\Delta} \tau \equiv \frac{1}{\bar{|\tau|}N_{\text{samp}}}\Sigma_{i=1}^{N_{\text{samp}}}\left(\tau(t)-\bar{\tau}\right)$.
Normalizing the variance by the mean value changes the scaling of these fluctuations. 
A plateau for low pressure in $\bar{\Delta}N_0$ occurs at the transition previously identified. 
This plateau goes away as the pressure increases.
The pressure dependence in $Z(I)$ causes a $\bar{\Delta}Z(I)$ pressure dependence, as is expected.  
However, the slope of $\bar{\Delta}Z(I)$ changes sign twice as $I$ increases.
The quasi-static transition is present for $\bar{\Delta}a_c$, although is less pronounced.

A $\sqrt{N}$ dependence in $\Delta \tau,\Delta N_1$ and $\Delta N_0$ for flows slower than the transition in $P^{-0.5}I$ is shown in Figure \ref{fig:varnorm}. 
This indicates an additional sensitivity to system size near arrest in the quasi-static regime for $\tau, N_1$ and $N_0$.
The fluctuation scaling in the Supplemental Material~\cite{Note1} shows collapse of $\bar{\Delta} \tau$ when $P^{-0.5}I$ is scaled by $\sqrt{N}$ dependence in $\bar{\Delta} \tau$ for flows slower than the transition.  
The additional system-size dependence in the fluctuations $\Delta N_1$ may influence the average first normal stress difference $N_1/P$, in Figure~\ref{fig:N0N1}b.
The spread of $N_1/P(I^2)$ values below the minimum $I^2<10^{-4}$ increases as the system size decreases.

\begin{figure}[]
\centering	
\includegraphics[width=0.8\columnwidth]{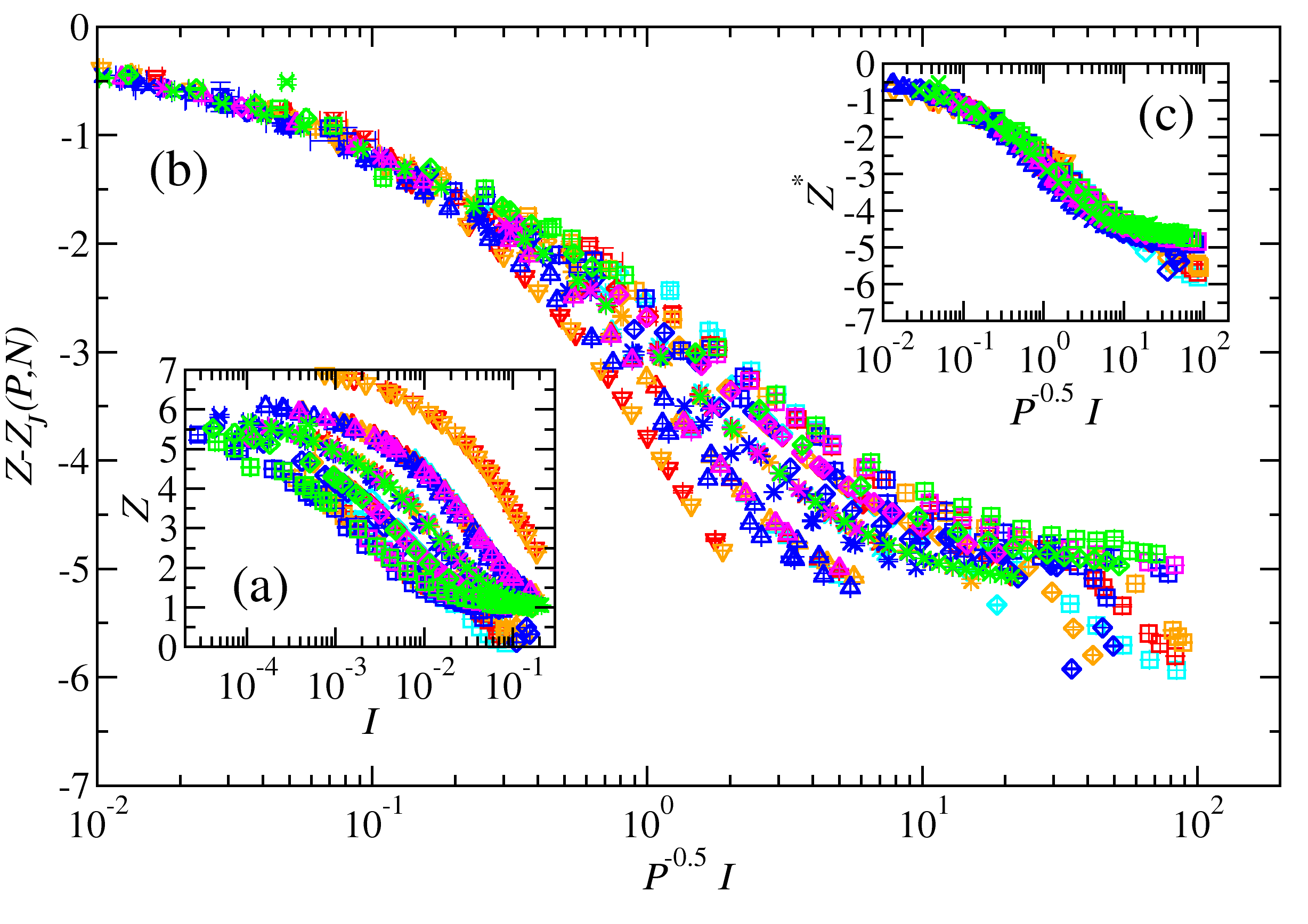}
\caption{(a) The coordination number $Z$ as a function of the inertial number $I$. (b) The distance from with the isotropic compression jamming coordination number $Z(I,P)-Z_J(P,N)$ as a function of the pressure-scaled inertial number $P^{-0.5} I$. (c) The hard-sphere limit coordination number $Z^*=Z(I,P)-Z_J(P,N)-\hat{Z}_{\text{soft}}(P)+\hat{Z}_{\text{hard}}$ as a function of the pressure-scaled inertial number $P^{-0.5} I$ . Colors and symbols are the same as in Fig.~\ref{fig:muphi-I}.}
\label{fig:z-I}
\end{figure}

Most of the properties presented have $P$- and $N$-dependent variances and $P$- and $N$-independent averages. 
Yet, the coordination number exhibits the opposite behavior; the variance $\Delta Z(I)$ is pressure-independent and the average $Z(I)$ is pressure-dependent, as shown in Fig.~\ref{fig:z-I}a.
As pressure increases $Z$ shifts to higher $I$, even though the volume fraction $\phi$ is pressure-independent
Therefore the inertial number is insufficient to capture the coordination number behavior.
The average $Z$ has the same $P$-dependence as the normalized $\Delta \tau$.
The collapse in Fig.~\ref{fig:z-I}b when plotted as a function of $P^{-0.5}I$ demonstrates the added $P$-dependence.
Figure~\ref{fig:z-I}b shows the distance of $Z(I)$ from the pressure and system size dependent jamming coordination number $Z_J(P,N)$, where $Z_J(P,N)$ was calculated using an isotropic pressure-controlled, zero shear protocol~\cite{Santos2020}. 
The average packing coordination number scales as $Z_J\sim P^{-0.5}$, as seen previously~\cite{OHern2003}.

In quasi-static flows, $Z(I,P) - Z_J(P,N)\equiv\hat{Z}(I,P)$ depends only on $P^{-0.5}I$. 
Faster flows ($P^{-0.5}I>0.2$) depend on pressure.
The coordination number $Z(I,P)$ increases with pressure, but the distance from the jamming coordination number $\hat{Z}(I,P)$ decreases at a faster rate as $P^{-0.5}I$ increases for higher pressures.
The inertial flow behavior of $\hat{Z}(I,P)$ is more complex.
Chivalo \textit{et al.}~\cite{Chialvo2012} observed a similar pressure effect on the shear stress ratio in volume-controlled simulations of inertial flows, and was attributed to the softness of high pressure systems.
A similar approach is used here, by fitting a power law 
\begin{equation}
Z(I,P)-Z_J(P,N)\equiv \hat{Z}(I,P) = A_{\hat{Z}} \left(P^{-0.5}I\right)^{\alpha_{\hat{Z}}}
\end{equation} 
\noindent to the pressure-shifted coordination number distance from jamming.
The fitting parameters, $A_{\hat{Z}}$ and $\alpha_{\hat{Z}}$ are shown in the Supplemental Material~\cite{Note1}.
Both fitting parameters are relatively insensitive to the system size $N$, however they both have a pressure dependence.
The power-law fitting parameters are well fit to the square of the pressure $\alpha_{\hat{Z}} = \alpha_{\hat{Z}, c} + B_{\alpha_{\hat{Z}}}\sqrt{P}$ and $A_{\hat{Z}}= A_{\hat{Z}, c} + B_{A_{\hat{Z}}}\sqrt{P}$.
The hard-sphere behavior, $P\to 0$, is extrapolated from the fitting parameters to be
\begin{equation}
\hat{Z}_{\text{hard}}(P^{-0.5}I) = -2.592\left(P^{-0.5}I\right)^{0.2861}
\end{equation}
\noindent, see the Supplemental Material~\cite{Note1} for more information.
Figure~\ref{fig:z-I}c shows the distance of pressure-dependent soft-sphere behavior from the hard-sphere behavior, which is subtracted from the simulated data
\begin{equation} 
Z^*=\hat{Z}(I,P) - \hat{Z}_{\text{soft}}(I,P) + \hat{Z}_{\text{hard}}(P^{-0.5}I)
\end{equation}.

\section{Conclusion}\label{sec:conclusion}
Power-law scalings of the fluctuations in various kinematic, mechanical and microstructural properties with pressure and system size were quantified in dense flows of dry frictionless granular materials.
Between 3x10$^2$ and 10$^5$ 3D spherical particles were flowed under simple shear in stress-controlled discrete-element, particle-based simulations.
Power law relationships between the mean steady state values of the stress ratio $\mu$, volume fraction $\phi$ and second normal stress $N_0/P$ as function of the inertial number $I$ were demonstrated.
Unlike those properties, the first normal stress $N_1/P$ shows a non-monotonic variation with $I$ which cannot be captured by the typical power law fits and requires large system sizes to measure with certainty.
Power-law fits are consistent across $N$ and $P$, for large systems sizes $N\ge1000$ and low pressures  $P<10^{-3}$.
A wide range of $I$ is needed to attain reliable fits that are comparable across $N$ and $P$, which requires many simulations.

The lower range of $I$ is limited by arrest near the critical shear stress, especially in the stress-controlled method used, where the arrest occurs stochastically.
The arrest stress ratio, along with fitted critical stress ratio, show strong system-size dependence at lower pressures. 
Further more the arrest $\mu_{\text{c,arrest}}$ and fitted critical $\mu_{\text{c,fit}}$ stress ratios have opposite dependencies to system size.
The $N$ dependence of $\mu_{\text{c,fit}}$ in the stress-controlled simulations are the opposite of what has been seen in strain-controlled simulations, because arresting flows limit the range of $I$ available giving more freedom for the fitting procedure. 
A more detailed measurement of the flow-to-arrest and arrest-to-flow transition for different system sizes is a subject of future study.

System size and pressure effects on fluctuations in steady flowing properties are considerably more pronounced than in their average properties.
The averages of $\mu, \phi, N_0/P$ and $N_1/P$ have the same relationship to inertial number regardless of $N$ and $P$.  
The fluctuations of each property, however, vary differently with $P$.
All fluctuations scale with $\sqrt{N}$, consistent with the central limit theorem.
The impact of $P$ on fluctuations differs in that they are either independent of $P$ ($\Delta\phi$, $\Delta Z$) or scale with $P$ (exponent magnitudes less than 1 for $\Delta\phi$, $\Delta a_c$  and $\Delta Z$) or scale strongly with $P$ (exponent magnitudes greater than 1 for $\Delta\tau, \Delta N_1$ and $\Delta N_0$).
Furthermore, $P$ impacted the fluctuation scaling with respect to inertial number for many properties.
Specifically, $\Delta\tau, \Delta a_c, \Delta N_1$ and $\Delta N_0$ do not collapse with $I$, but with $P^{-0.5}I$ and have pressure-dependent transition at $P^{-0.5}I\simeq0.2$.
The slope changes sign for $P^{-0.5}I\simeq0.2$, $\Delta\tau(I) \Delta N_1(I)$ and $\Delta N_0(I)$ at that transition.
Power-law fits to those fluctuations above and below the transition were presented.
The transition in the variance of shear stress is a potential quantitative measure of the boundary between quasi-static and inertial flow regimes.

Interestingly, fluctuations in the coordination number are not pressure-sensitive, while the average coordination number is sensitive to pressure.
The average $Z(I)$ is pressure-dependent, unlike the other properties presented, and requires $I$ to be scaled by $P^{-0.5}I$, as was seen in $\Delta\tau, \Delta a_c, \Delta N_1$ and $\Delta N_0$.
The pressure dependence in the coordination number flow behavior in the quasi-static flow regime is captured by $Z-Z_J=\hat{Z}(P^{-0.5}I)$.
In the inertial regime, there is an additional pressure dependence stemming from effective particle softness in faster flows. 
We take the limiting behavior of $\hat{Z}(P^{-0.5}I)$ to define hard-sphere limit behavior, and the distance from that hard-sphere limit accounts for the additional pressure dependence of the coordination number in the inertial regime.

This is the first comprehensive study to quantify not just the steady state values of important microstructural metrics such as coordination and fabric anisotropy in dense granular flows, but also the fluctuations of these properties and their system size scaling. 
These results could greatly contribute towards the development of microstructure-aware constitutive models for granular flows, particularly those that include the role of fluctuations, as was demonstrated previously for dilute granular gases~\cite{VanNoije1997}.
Such a model could be very useful for small confined systems where fluctuations have a crucial role.
Ongoing work includes the effect of sliding, rolling and twisting friction on the flow behavior. 

\section{Acknowledgments}
This work was supported by the Sandia Laboratory Directed Research and Development Program.
This work was performed, in part, at the Center for Integrated Nanotechnologies, an Office of Science User Facility operated for the U.S. Department of Energy (DOE) Office of Science. 
I. S. acknowledges support from the U.S. Department of Energy, Office of Science, Office of Advanced Scientific Computing Research, Applied Mathematics Program under contract No. DE-AC02-05CH11231. 
Sandia National Laboratories is a multi-mission laboratory managed and operated by National Technology and Engineering Solutions of Sandia, LLC., a wholly owned subsidiary of Honeywell International, Inc., for the U.S. DOE’s National Nuclear Security Administration under contract DE-NA-0003525. 
The views expressed in the article do not necessarily represent the views of the U.S. DOE or the United States Government.

\bibliographystyle{apsrev4-2}
\bibliography{FrictionlessGranularSystemSize.bib}
\end{document}


\preprint{APS/123-QED}
\title{Supplement Material for: Fluctuations and power-law scaling of dry, frictionless granular rheology near the hard-particle limit}
\author{A. P. Santos}
\email{asanto@sandia.gov}
\affiliation{Sandia National Laboratories, Albuquerque, NM 87185, USA}
\author{Ishan Srivastava}
\affiliation{Center for Computational Sciences and Engineering, Lawrence Berkeley National Laboratory, Berkeley, CA 94720, USA}
\author{Leonardo E. Silbert}
\affiliation{School of Math, Science and Engineering, Central New Mexico Community College, Albuquerque, NM 87106, USA}
\author{Jeremy B. Lechman}
\affiliation{Sandia National Laboratories, Albuquerque, NM 87185, USA}
\author{Gary S. Grest}
\affiliation{Sandia National Laboratories, Albuquerque, NM 87185, USA}
\date{\today}

\maketitle

\begin{figure}
	\includegraphics[width=0.5\linewidth]{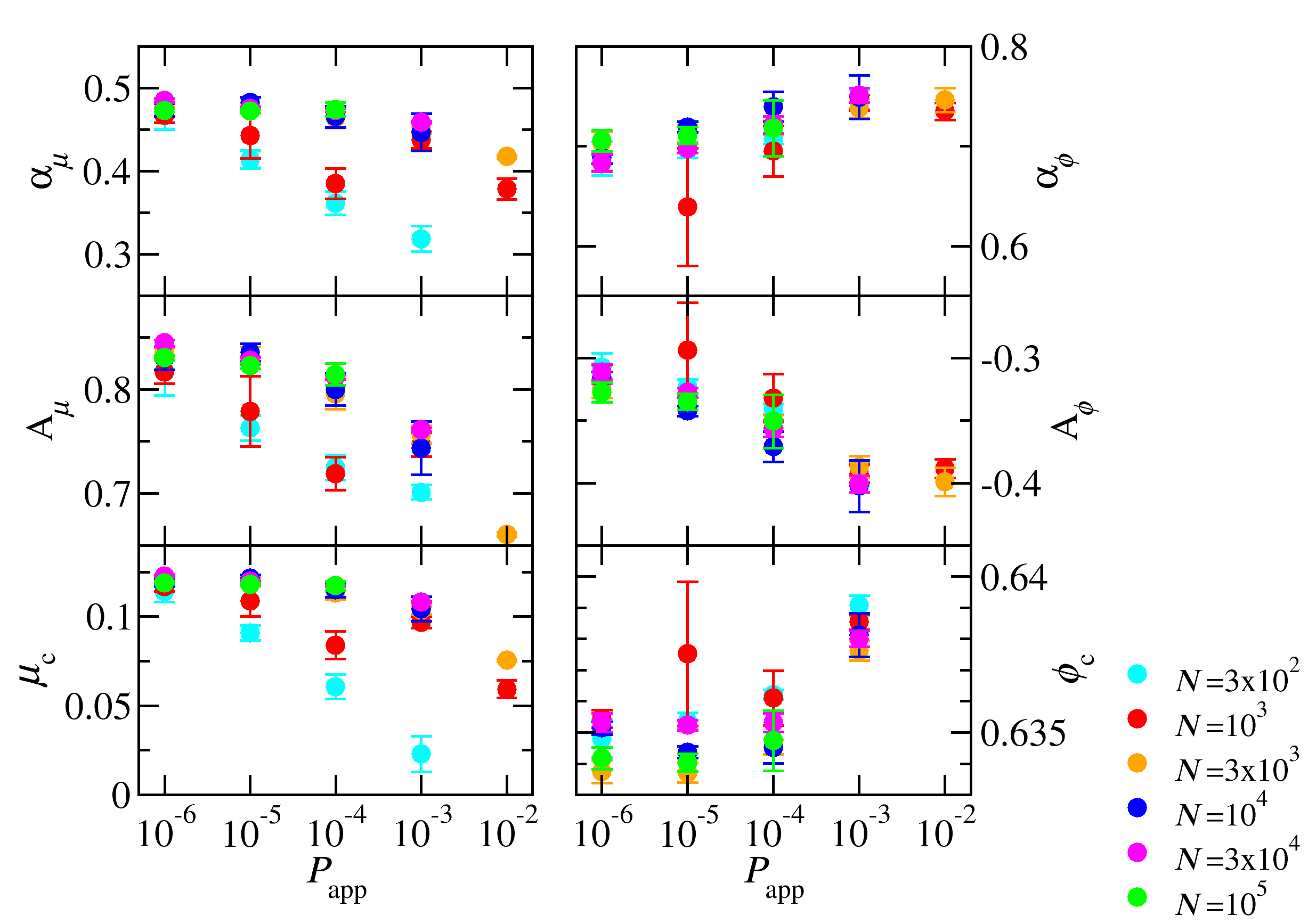}
	\caption{Stress ratio $\mu(I)$ (left three panels) and volume fraction $\phi(I)$ (right three panels) are fit to Eqs.~6 and~7 in the main text, respectively, to get exponents $\alpha_{\mu},\alpha_{\phi}$, pre-factors $A_{\mu},A_{\phi}$ and critical values $\mu_c,\phi_c$ as a function of pressure $P$. Different system sizes $N$ are shown as different colors $P=10^{-7}$ (circles), $P=10^{-6}$ (squares),  $P=10^{-5}$ (diamonds),  $P=10^{-4}$ (crosses), $P=10^{-3}$ (triangles).}
	\label{fig:muphifits-P}
\end{figure}

\begin{figure}[!htbp]
	\centering	
	\includegraphics[width=0.5\columnwidth]{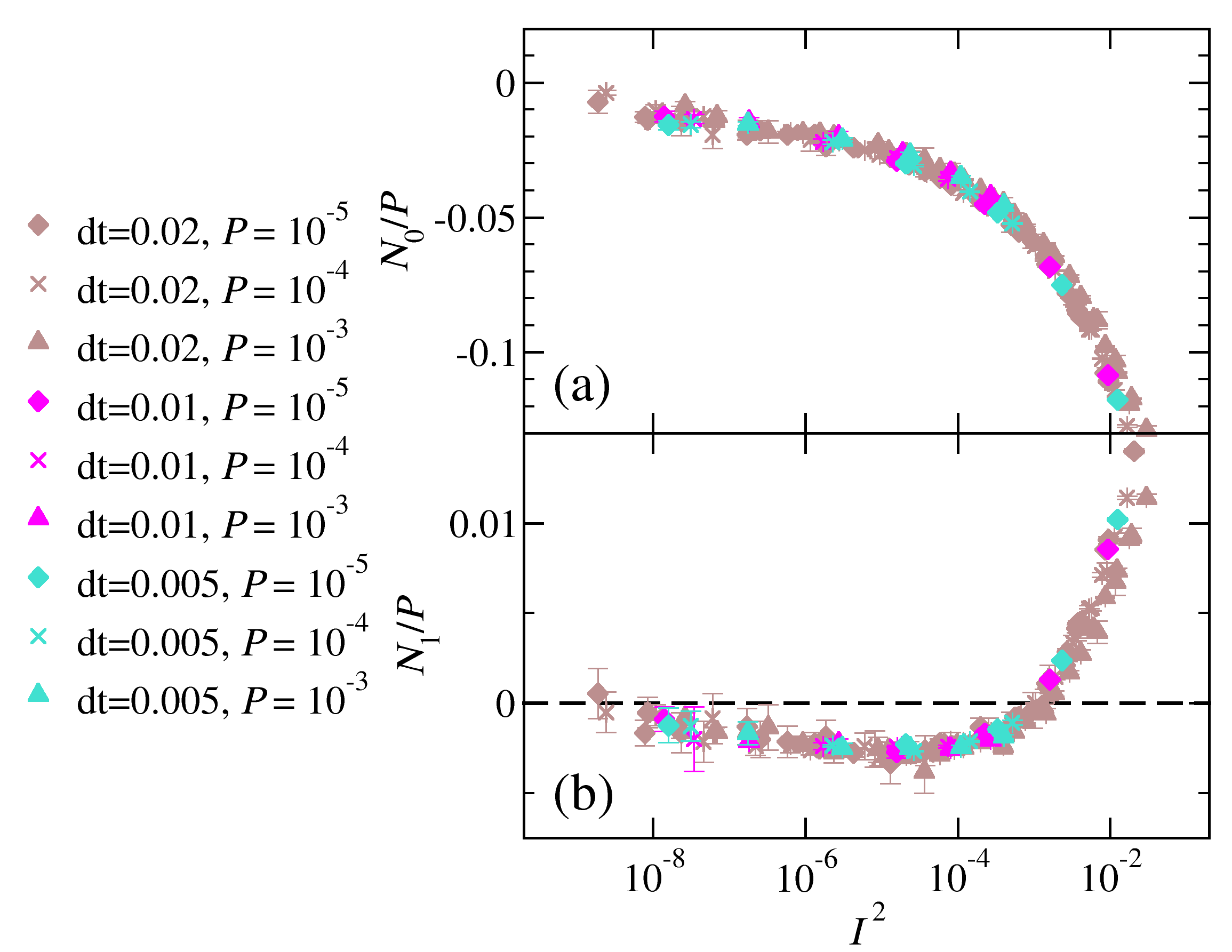}
	\caption{(a) Second $N_0/P$ and (b) first $N_1/P$ normal stress differences normalized by pressure (see Eqs.~4 and~5 in the main text) as a function of $I^2$ for different time steps dt and pressures $P$.  All data is for $N=10^4$.}
	\label{fig:N1-I}
\end{figure}

\begin{figure}
	\includegraphics[width=\linewidth]{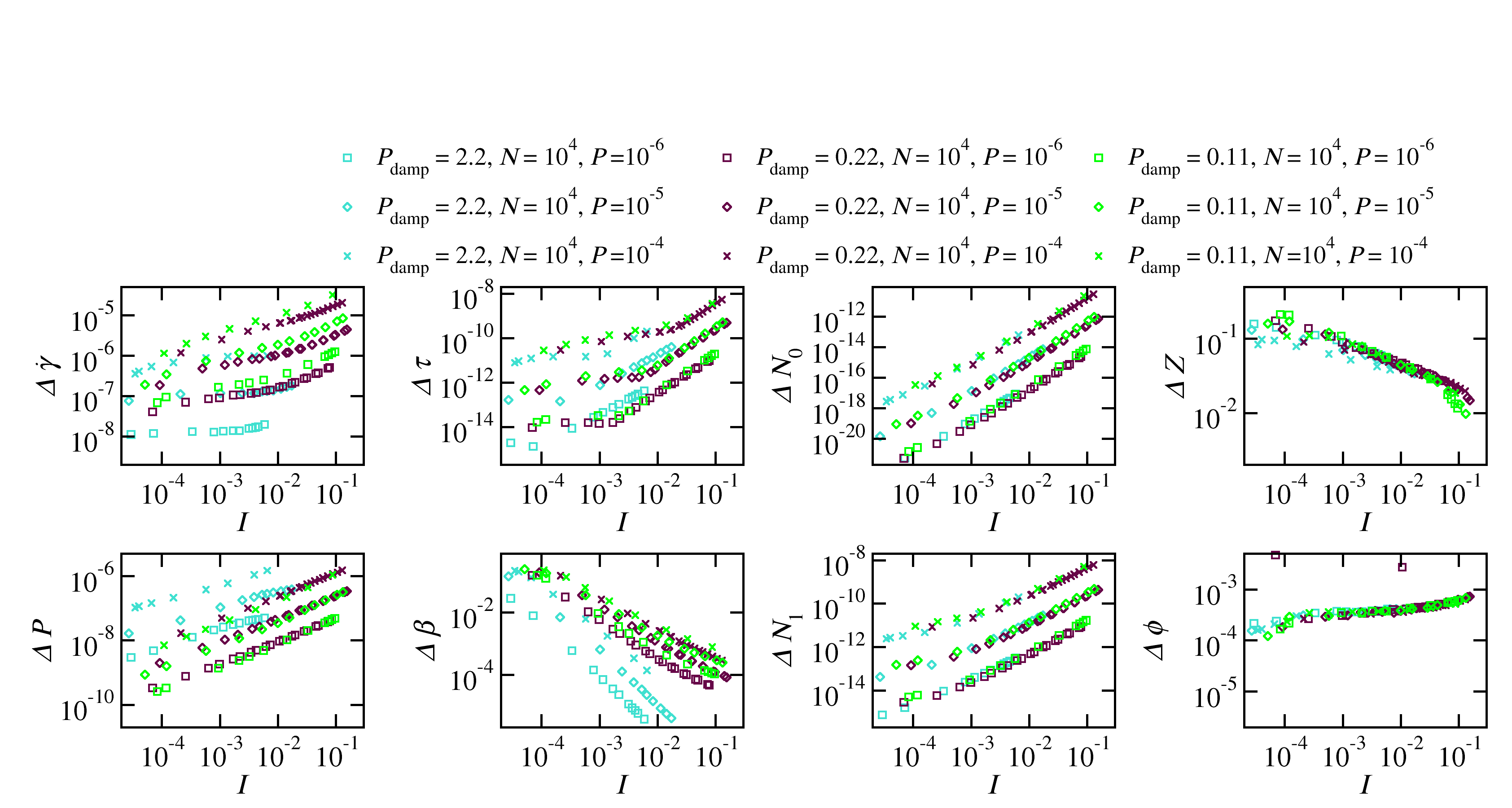}
	\caption{The variance $\Delta$ of stead-state fluctuations for three pressures $P$ and three pressure dampings $P_{\text{damp}}$ of the following parameters as a function of the measured steady-state inertial number: pressure $P$, strain rate $\gamma$, flow parameter $\beta$, shear stress $\tau$, first normal stress $N_1$, second normal stress $N_2$, volume fraction $\phi$, and the coordination number $Z$.	}
	\label{fig:varpdamp}
\end{figure}

\begin{figure}
	\includegraphics[width=\linewidth]{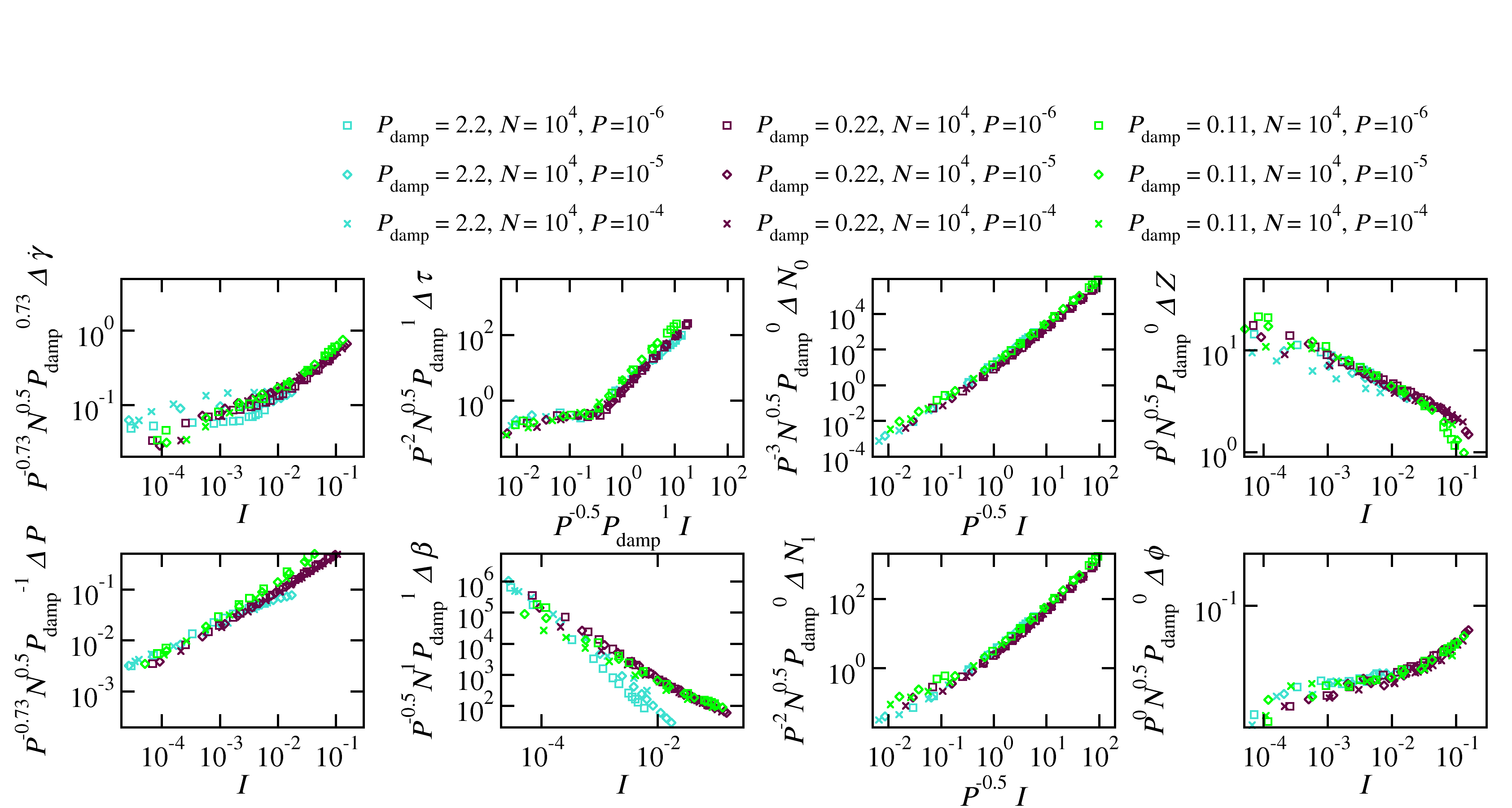}
	\caption{Figure~\ref{fig:varpdamp} where the variance $\Delta$ is normalized by $N^a$, $P^b$ and $P_{\text{damp}}^b$. This leads to a collapse as a function of the inertial number $I$, except for the stress values, $\tau$, $N_1$ and $N_2$, which are normalized by $P^c$}
	\label{fig:varnormpdamp}
\end{figure}

\begin{table}
	\begin{tabular}{ c|ccc|cccc } 
		\hline
		\hline
		&  & exponent &  &  &  &  &  \\ 
		& $a$ & $b$ & $c$ & $d$ & $e^*$ &  $d^+$ & $e^+$  \\ 
		\hline
		$P$ 					& 0.5 & 0.3 & 1 & 0.686$\pm$0.003 & 1.14  & &  \\ 
		$\dot{\gamma}$ & 0.5 & -0.25 & 1 &0.381$\pm$0.008 & 0.92 & &  \\ 
		$a_c$ 			& 0.5 & 0.07 & -0.5 & -1.10$\pm$0.01 & -0.05 & & \\ 
		$\tau$ 		   & 0.615 & 0.125 & -0.5 & 0.46$\pm$0.03 & 1.54 & 1.626$\pm$0.007 & 0.37  \\ 
		$N_1$ 				& 0.5 & 0.0625 & -0.5 & 1.18$\pm$0.01& 0.82 & &  \\ 
		$N_2$ 				& 0.5 & -1 & -0.5 & 2.22$\pm$0.01& 0.78 & &  \\ 
		$\phi$ 				& 0.5 & 0 & 1 & 0.12$\pm$0.02 & 0.06 & 0.58$\pm$0.07  & 0.29  \\ 
		$Z$ 				& 0.5 & 0.167 & -0.5 & -0.28$\pm$0.01 & -0.14 & & \\
		\hline
	\end{tabular}
	\caption{Exponent values used to scale and normalize the the relative variance of various properties run at different $N$ and applied pressure as a function of inertial number.
		\newline$^*$The relative variance-pressure exponent $e=d(c-1/2)-b$. 
		\newline$ ^+$Property relative variance with different fits at high strain rates.}
	\label{tab:exponents}
\end{table}

\begin{figure}
	\includegraphics[width=\linewidth]{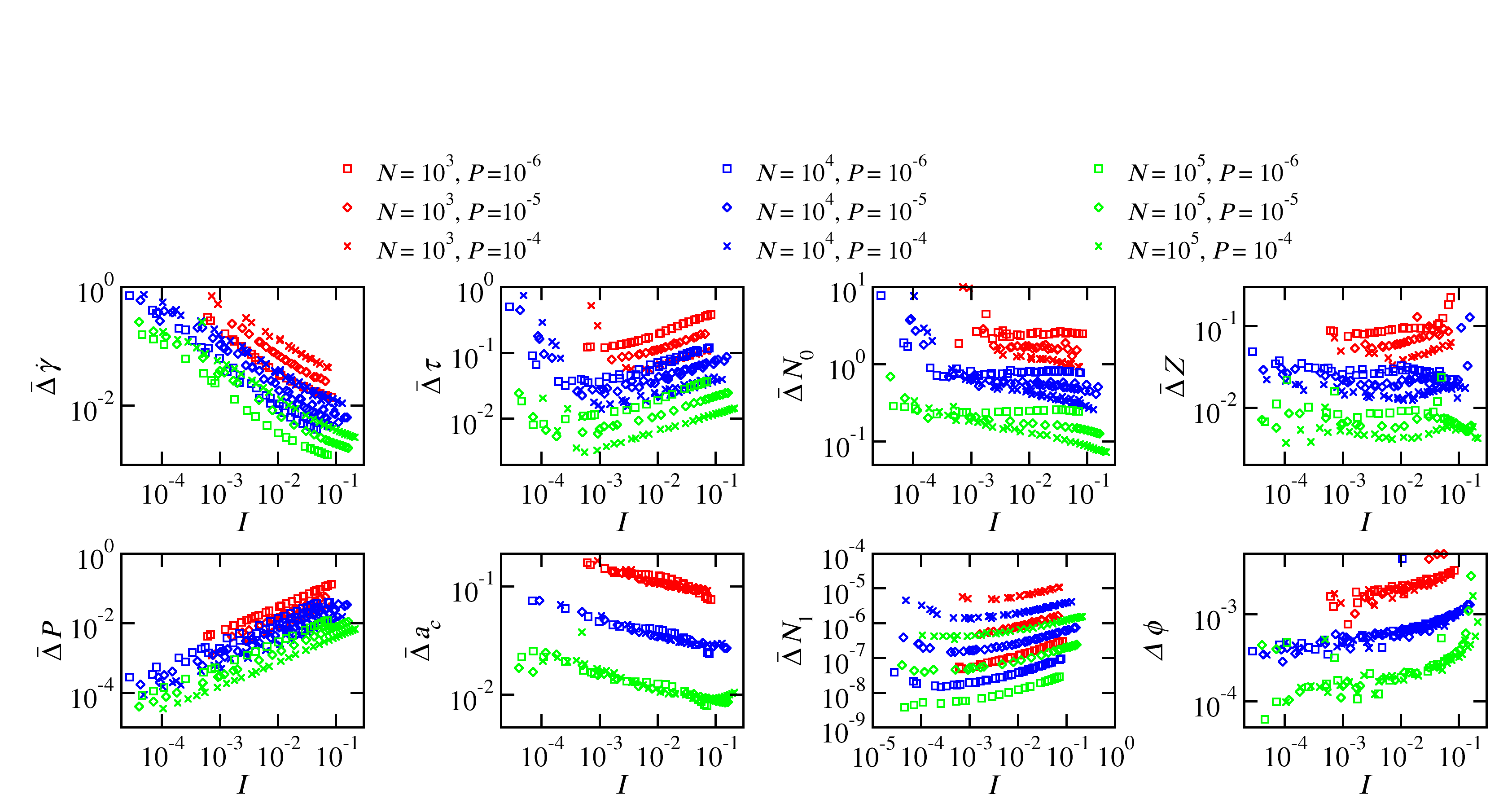}
	\caption{The relativ variance, not normalized by the mean $\bar{\Delta}$ of stead-state fluctuations for three pressures $P$ and three pressure dampings $P_{\text{damp}}$ of the following parameters as a function of the measured steady-state inertial number: pressure $P$, strain rate $\gamma$, flow parameter $\beta$, shear stress $\tau$, first normal stress $N_1$, second normal stress $N_2$, volume fraction $\phi$, and the coordination number $Z$. }
	\label{fig:relativevar}
\end{figure}

\begin{figure}
	\includegraphics[width=\linewidth]{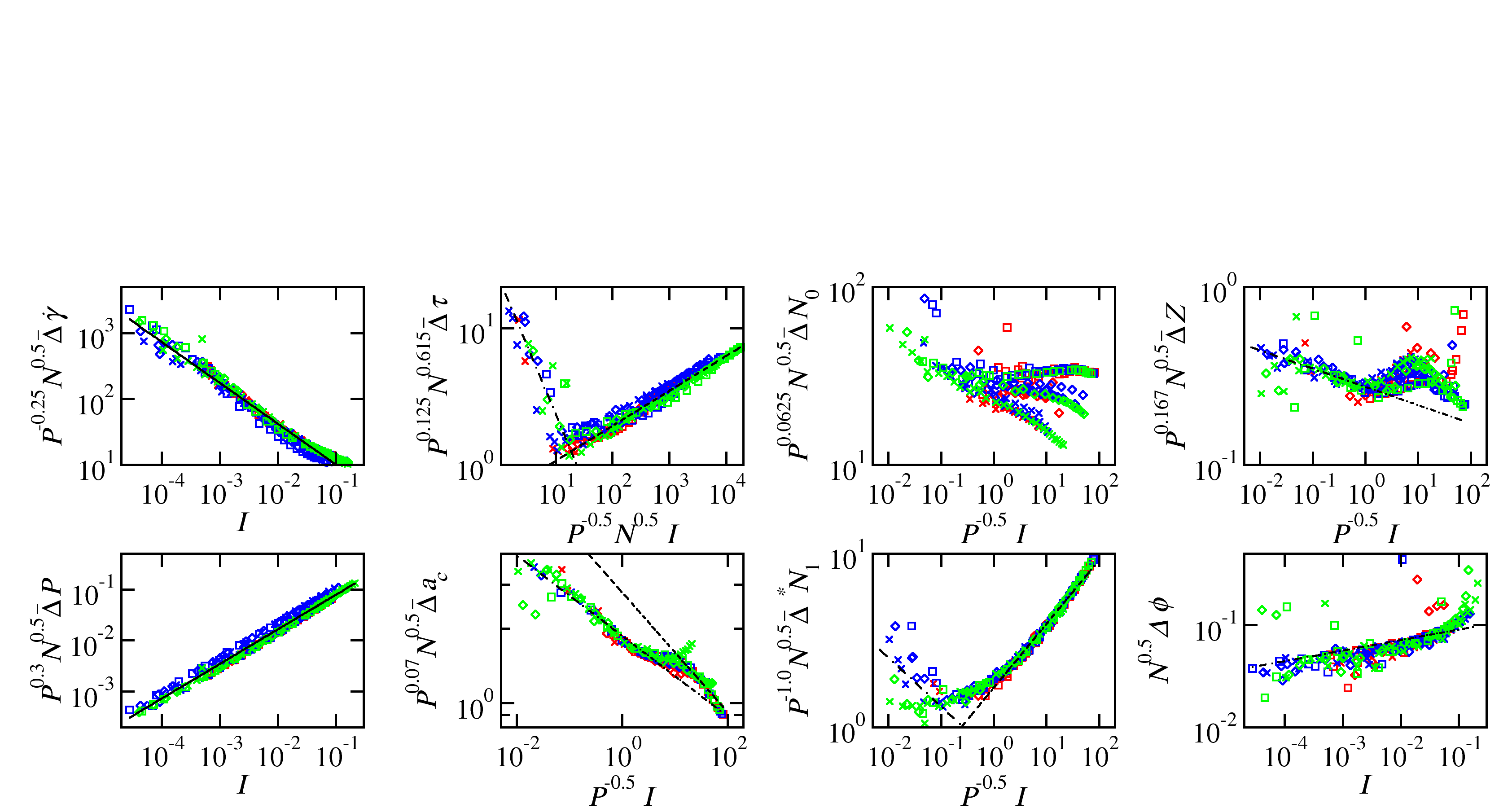}
	\caption{Figure~\ref{fig:relativevar} where the relative variance $\bar{\Delta}$ is normalized by $N^a$, $P^b$ and $P_{\text{damp}}^b$. This leads to a collapse as a function of the inertial number $I$, except for the stress values, $\tau$, $N_1$ and $N_2$, which are normalized by $P^c$.}
	\label{fig:relativevarnorm}
\end{figure}

\begin{figure}
	\includegraphics[width=0.5\linewidth]{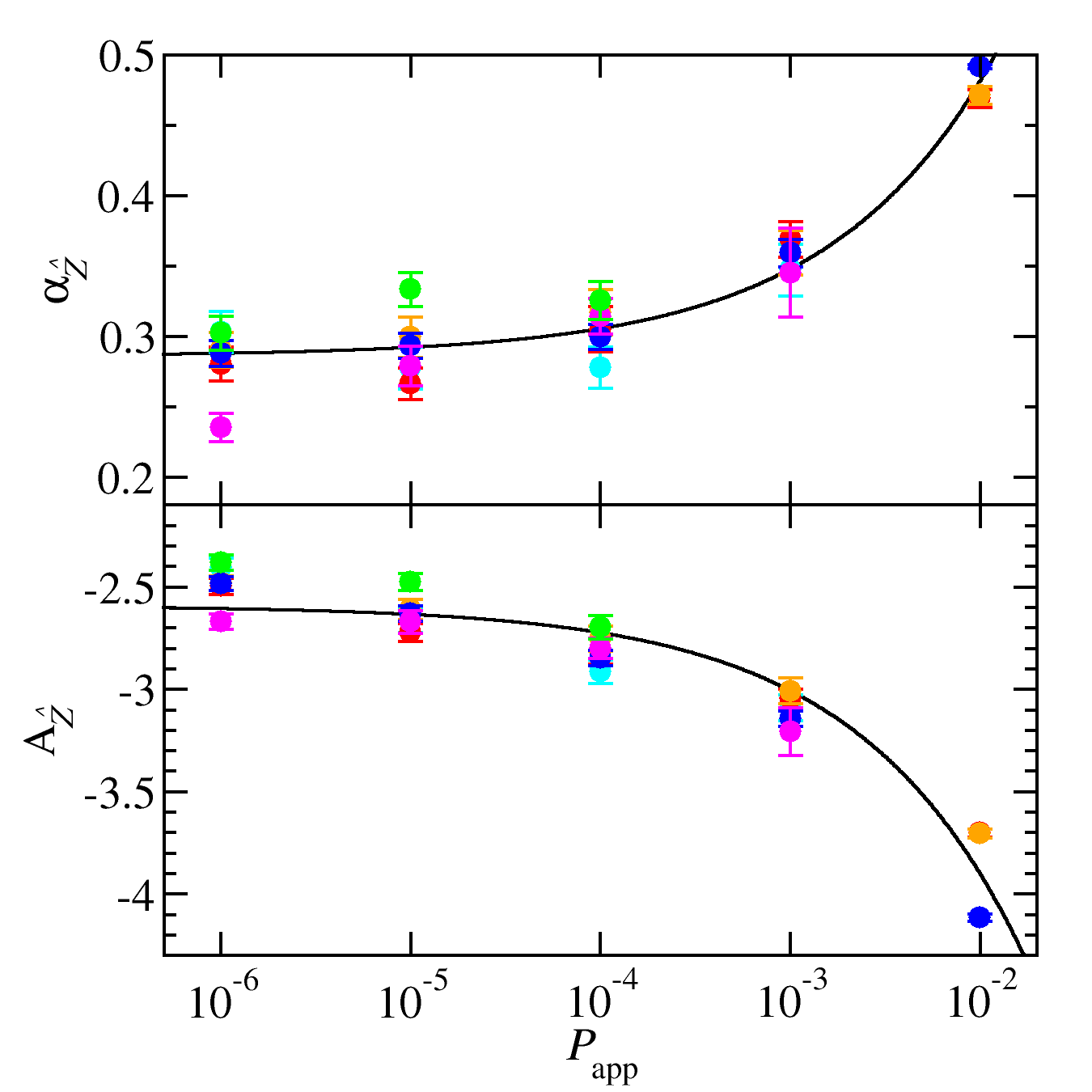}
	\caption{Fit parameters to $Z-Z_J(P,N)$, shown in Fig.~8 of the main text, of the form $Z-Z_J(P,N)\equiv\hat{Z} = A_{\hat{Z}}\left(I/\sqrt{P}\right)^{\alpha_{\hat{Z}}}$. 
	Different colored symbols represent different system sizes $N=$ 300 (cyan), 1,000 (red), 3,000 (orange), 10,000 (blue), 30,000 (magenta), 100,000 (green). 
	The solid black curve is a fit to the fit parameters themselves of the form $\alpha_{\hat{Z}} = \alpha_{\hat{Z}, c} + B_{\alpha_{\hat{Z}}}\sqrt{P}$ and $A_{\hat{Z}} = A_{\hat{Z}, c} + B_{A_{\hat{Z}}}\sqrt{P}$. 
	Linear regression leads to fit parameters that define the black curve $\alpha_{\hat{Z}, c} = 0.2861 \pm 0.0010$, $B_{\alpha_{\hat{Z}}} = 1.95 \pm 0.03$, $A_{\hat{Z}, c} = -2.592 \pm 0.007$ and $B_{A_{\hat{Z}}} = -13.1 \pm 0.2$ with $R^2=0.91$. }
	\label{fig:relativevarnorm}
\end{figure}